\documentclass[a4paper,english,prd,showkeys]{revtex4-2}
\usepackage{color}
\usepackage{array}
\usepackage{multirow}
\usepackage{amsmath}
\usepackage{amssymb}
\usepackage{graphicx}
\usepackage{wasysym}



\usepackage{babel}
\begin{document}

\title{Vacuum Birefringence in a Supercritical Magnetic Field and a Subcritical
Electric Field}

\author{Chul Min Kim}
\email{chulmin@gist.ac.kr}
\affiliation{Center for Relativistic Laser Science, Institute for Basic Science,
Gwangju 61005, Korea}
\affiliation{Advanced Photonics Research Institute, Gwnagju Institute of Science
and Technology, Gwangju 61005, Korea}
\author{Sang Pyo Kim}
\email{sangkim@kunsan.ac.kr}
\affiliation{Department of Physics, Kunsan National University, Kunsan 54150, Korea}
\date{\today}

\begin{abstract}
Recent  ultra-intense lasers of subcritical fields and proposed observations of the x-rays polarization from highly magnetized neutron stars of supercritical fields have attracted attention to vacuum birefringence, a unique feature of nonlinear electrodynamics.
We propose a formulation of vacuum birefringence that incorporates
the effects of the weaker electric field added to the extremely strong
magnetic field. To do so, we first derive a closed analytical expression for the one-loop
effective Lagrangian for the combined magnetic and electric fields
by using an explicit formula of the one-loop effective Lagrangian
for an arbitrarily strong magnetic field. 
We then employ the expression to derive the polarization and magnetization of the vacuum,
from which the permittivity and permeability for weak probe fields
are obtained. Finally, we find the refractive indices and the associated polarization
vectors for the case of parallel magnetic and electric fields. The proposed formulation predicts that an electric field along
the magnetic field reduces the birefringence and rotates the polarization
vectors. Such effects should be taken into account for accurate polarimetry
of the x-rays from magnetized neutron stars, which will
prove the fundamental aspect of the strong field quantum electrodynamics (QED) and explore the
extreme fields of astrophysical bodies.
\end{abstract}

\maketitle

\section{Introduction}

A background electromagnetic field polarizes the Dirac vacuum and
produces charged particles-antiparticle pairs. Heisenberg-Euler and
Weisskopf found the one-loop effective Lagrangian in a constant electromagnetic
field \citep{Heisenberg1936Folgerungen,Weisskopf1936Ueber}, and Schwinger
obtained the effective Lagrangian in the proper-time integral by integrating
out the fermion or scalar boson coupled to the electromagnetic field
within quantum electrodynamics (QED) \citep{Schwinger1951Gauge}.
The imaginary part of the proper-time integral, contributed by simple poles, gives the loss of the vacuum persistence due to pair production. In fact, a pure electric field or electric field parallel
to a magnetic field in a proper Lorentz frame produces electron-positron
pairs from the Dirac sea via quantum tunneling through the tilted mass gap.
The correct electromagnetic theory should be described by the effective
action consisting of the Maxwell action and the loop corrections due
to strong electromagnetic fields. Hence, a probe photon propagates
through a polarized vacuum due to the effective action, and prominent
phenomena of vacuum polarization can occur such as vacuum birefringence,
photon splitting, photon-photon scattering, etc \citep{Melrose2013Quantum}.

The pair production, called the Schwinger effect, is a nonperturbative
effect of quantum field theory. The electron-positron pairs can be
efficiently produced when the electric field is comparable to the
critical field $E_{\mathrm{cr}}=m^{2}c^{3}/e\hbar=1.3\times10^{16}\,\mathrm{V/cm}$
since the pair production rate per a unit four Compton volume is given
by the Boltzmann factor of which exponent is given as the negative
of the ratio of the critical field to the electric field. The magnetic
field of the critical strength, $B_{\mathrm{cr}}=m^{2}c^{3}/e\hbar=4.4\times10^{13}\,\mathrm{G}$,
makes the lowest Landau energy equal to the rest mass of the electron.
The critical strengths $E_{\mathrm{cr}}$ and $B_{\mathrm{cr}}$ are
called the Schwinger fields. The effect of vacuum polarization and
the Schwinger pair production can be measured when the electromagnetic
fields are comparable to or higher than the critical field. Thus,
the physics in strong electromagnetic fields that is governed by the
effective action drastically differs from the physics in weak fields
that obeys the Maxwell theory.

In experiments, the Schwinger pair production is still very difficult
to realize because no terrestrial mean provides an electric field
comparable to the Schwinger field. In spite of the recent progress
in high-intensity lasers based on chirped pulsed amplification (CPA)
technique, the current highest laser intensity is $1.1\times10^{23}\,\mathrm{W/cm^{2}}$,
achieved by CoReLS \citep{Yoon2021Realization}, of which field strength
is still lower than the critical field by three orders. Several laser
facilities are being constructed for higher intensities, but the target
fields strengths are still lower by order one or two \citep{Danson2019Petawatt}.

In contrast, the effects of vacuum polarization such as photon-photon
scattering have been experimentally investigated. The Delbr\"uck scattering, in which a photon is scattered by a Coulomb field, was observed with MeV photons \citep{Moreh1973Delbruck,Rullhusen1983Giant}. The photon splitting, in which a photon is split into two  by a Coulomb field was also observed \citep{Jarlskog1973Measurement}. These observations were enabled by the strong nuclear Coulomb field. The photon-photon scattering without a Coulomb field is more difficult to realize but was recently evidenced from heavy ion collision experiments: the ATLAS
experiment \citep{2013Observing,ATLASCollaboration2017Evidence} and the CMS experiment  \citep{Sirunyan2019Evidence}. 

The photon-photon scattering can also occur under a magnetic field. The Delbr\"uck scattering by a magnetic field leads to the vacuum birefringence, and, as a consequence, the vacuum under a strong magnetic field can act as a birefringent medium to low-energy photons. Compared to the Schwinger pair production
and the photon-photon scattering by the nuclear Coulomb field, the vacuum birefringence has the
advantage of accumulating the effect over a macroscopic length scale.
To realize the vacuum birefringence, the PVLAS project uses the magnetic
field from a strong permanent magnet as the background field and optical laser photons as the probe photons \citep{Valle2016PVLAS}. Also, it was proposed to use the field from an ultra-intense laser as the background field and the strong x-rays from an x-ray free electron laser as the probe photons \citep{Karbstein2021Vacuum,Shen2018Exploring}. This proposal relies on the state-of-arts scientific technologies such as ultra-intense lasers \citep{Danson2019Petawatt}, x-ray free electron lasers \citep{Pellegrini2016X}, and ultra-high-precision x-ray polarimetry \citep{Schmitt2021Disentangling}. Albeit challenging, the goal of the PVLAS project and the laser-based proposal is limited to the vacuum birefringence in subcritical fields.

On the other hand, highly magnetized neutron stars have magnetic fields
comparable to the critical field, and particularly magnetars have
magnetic fields stronger than the critical field \citep{Vasisht1997discovery,Olausen2014McGILL,Kaspi2017Magnetars}.
The dipole model for pulsars and highly magnetized neutron stars provides
strong dipole magnetic fields and weak induced electric fields \citep{Goldreich1969Pulsar}.
Therefore, it will be interesting to study the QED vacuum polarization
effect in such supercritical magnetic fields and subcritical electric
fields, which can provide a diagnostics for strong electromagnetic
fields of neutron stars \citep{Enoto2019Observational}. Recently,
the observation of the x-rays from highly magnetized neutron stars
by using space telescopes has been proposed \citep{Santangelo2019Physics,Wadiasingh2019Magnetars}.
Furthermore, the scale of field variation is the radius of the neutron
star so that the one-loop effective action can be accurately given
by the Heisenberg-Euler and Schwinger action.

To explicitly express the vacuum birefringence in supercritical magnetic
fields, a closed\textcolor{red}{{} }analytic expression of one-loop
effective action is more convenient than the proper-time integral
expression obtained by Heisenberg-Euler and Schwinger. Dittrich employed
the dimensional regularization method to perform the proper-time integral
in terms of the Hurwitz zeta-function and logarithmic functions in
either a pure magnetic field or an electric field perpendicular to
the magnetic field \citep{Dittrich1976One,Dittrich1979Evaluation}.
In Ref.~\citep{Kim2019Quantum}, the in-out formalism that leads
to the proper-time integral also directly gives the closed analytic
expression for the one-loop effective Lagrangian in the same field
configuration, which is identical to the one by Dittrich. Furthermore,
the imaginary part of the one-loop Lagrangian in the closed form yields
the same result obtained by summing the residues of all simple poles
of the proper-time integral \citep{Kim2019Quantum}.

In this paper, we develop the method in \citep{Kim2021Magnetars}
to find the closed analytic expression for the one-loop effective
action in supercritical magnetic fields combined with subcritical
electric fields. Provided that the fields vary little over the Compton
length and time, one may use the Heisenberg-Euler and Schwinger one-loop
effective Lagrangian in the gauge\textendash{} and Lorentz\textendash invariant
form as a good approximation and can express the one-loop Lagrangian
as a power series of a small invariant quantity that becomes the electric
field in a Lorentz frame where magnetic and electric fields are parallel
to each other. Using the closed expression, we study the propagation
modes of a weak probe photon in a vacuum under such electromagnetic
fields. For this purpose, we find the permittivity and permeability
tensors and obtain the vacuum birefringence for a weak probe photon
in parallel electric and magnetic fields.

This paper is organized as follows. In Sec.~\ref{sec:Leff}, an explicit
expression of the one-loop effective Lagrangian is derived for the
vacuum under an arbitrarily strong magnetic field superposed with
a weaker electric field. The expression is given as a Taylor series
in a parameter that is essentially the ratio of the parallel component
of the electric field to the magnetic field. Then, in Sec.~\ref{sec:Response},
the series is used to obtain the permittivity and permeability tensors
for a weak low-frequency probe field. These tensors are used in Sec.~\ref{sec:VacBi}
to find the modes of the probe field (the refractive indices and associated
polarization vectors) for the configuration in which the electric
field is parallel to the magnetic field. Finally, a conclusion is
given, stressing that the presented formulae are necessary to analyze
the vacuum birefringence in the pulsar magnetosphere, in which extremely
strong magnetic field coexists with a weaker but non-negligible electric
field. The Lorentz-Heaviside units with $\hbar=c=1$ was used, in
which the fine structure constant is $\alpha=e^{2}/4\pi$ ($e$ the
elementary charge).

\section{One-loop Effective Lagrangian of the vacuum under a uniform electromagnetic
field\label{sec:Leff}}

\subsection{Invariant parameters and classification of uniform electromagnetic
fields}

When dealing with the effective Lagrangian of the vacuum in a constant
electromagnetic field, the following Lorentz- and gauge-invariant
parameters are convenient for analysis \citep{Dittrich2000Probing}:

\begin{equation}
a=\sqrt{\sqrt{F^{2}+G^{2}}+F},\quad b=\sqrt{\sqrt{F^{2}+G^{2}}-F},\label{ab_FG}
\end{equation}
where $F^{\mu\nu}$ and $F^{*\mu\nu}=\frac{1}{2}\varepsilon^{\mu\nu\alpha\beta}F_{\alpha\beta}$
($\varepsilon^{0123}=1$) are the field-strength tensor and its dual,
respectively \citep{Jackson1999Classical}. Then the Maxwell scalar
an and pseudoscalar are given as

\begin{equation}
F=\frac{1}{4}F^{\mu\nu}F_{\mu\nu}=\frac{1}{2}\left(\mathbf{B}^{2}-\mathbf{E}^{2}\right)=\frac{1}{2}\left(a^{2}-b^{2}\right),\quad G=\frac{1}{4}F^{\mu\nu}F_{\mu\nu}^{*}=-\mathbf{E}\cdot\mathbf{B}=\sigma ab,\label{FG_EB}
\end{equation}
where $\sigma$ denotes the sign of $G$. As will be shown later,
the formulation in terms of $a$ and $b$, instead of $F$ and $G$
or $\mathbf{E}$ and $\mathbf{B}$, has the advantage to facilitate
the expansion of the effective Lagrangian as a series of $b$.

\begin{figure}
\includegraphics[width=0.45\textwidth]{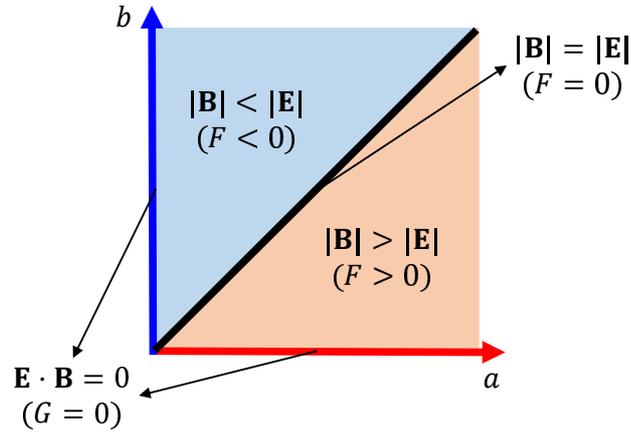}

\caption{Classification of constant electromagnetic fields in the $ab$-plane.
The diagonal line $b=a$ corresponds to the condition of equally strong
electric and magnetic fields, i.e., $|\mathbf{B}|=|\mathbf{E}|$ ($F=\left(\mathbf{B}^{2}-\mathbf{E}^{2}\right)/2=0$).
The horizontal and vertical axes correspond to the wrenchless condition,
i.e., $G=-\mathbf{E}\cdot\mathbf{B}=0$.}

\label{fig:classification}
\end{figure}

The parameters $a$ and $b$ can be used to classify the cases of
constant electromagnetic fields, as shown in Fig.~\ref{fig:classification}.
The sign of the Maxwell scalar $F$ determines which field is stronger
between the electric and magnetic fields, dividing the $ab$-plane
into two regions: the upper left where the electric field is stronger
and the lower right where the magnetic field is. The condition $G=0$
shrinks each region to its attached coordinate axis. The cases with
$G=0$ are called wrenchless, while the fields with $G\neq0$ are
said to have an electromagnetic wrench \citep{Melrose2013Quantum}.
In the wrenchless case, an appropriate Lorentz transformation can
remove the weaker field between the magnetic field and the electric
field \citep{Jackson1999Classical}. Thus, the $a$-axis ($b$-axis)
in Fig.~\ref{fig:classification} represents the condition of essentially
being under a magnetic (electric) field; of course, it includes the
case of a pure magnetic (electric) field. In studying the vacuum birefringence
of astrophysical relevance, the magnetic field is much stronger than
the electric field, and thus the region of $a\gg b$ in Fig.~\ref{fig:classification}
is of our interests.

\subsection{Integral expression of \textmd{\normalsize{}$\mathcal{L}^{(1)}(a,b)$}
and closed expression of \textmd{\normalsize{}$\mathcal{L}^{(1)}(a,0)$\label{subsec:Lab_La0} }}

The physics of the vacuum in intense electromagnetic fields has been
studied with the effective Lagrangian, which is obtained by integrating
out the matter field degrees of freedom in the complete Lagrangian
\citep{Schwartz2014Quantum}. In the effective Lagrangian, the term
other than the free-field part, the Maxwell theory ($\mathcal{L}^{(0)}(a,b)=(b^{2}-a^{2})/2$),
is responsible for the phenomena such as pair production, vacuum birefringence,
photon splitting, etc. As the term is dominantly contributed from
the one-loop \citep{Ritus1976Lagrangian,Gies2017addendum,Karbstein2019All}
at least for the magnetic field strengths of astrophysical relevance,
we consider the effective Lagrangian up to the one-loop:

\begin{equation}
\mathcal{L}_{\mathrm{eff}}(a,b)=\mathcal{L}^{(0)}(a,b)+\mathcal{L}^{(1)}(a,b)=\frac{b^{2}-a^{2}}{2}+\mathcal{L}^{(1)}(a,b).
\end{equation}
The one-loop contribution $\mathcal{L}^{(1)}(a,b)$ for the spinor
QED is given as a proper-time integral \citep{Heisenberg1936Folgerungen,Weisskopf1936Ueber,Schwinger1951Gauge}:

\begin{equation}
\mathcal{L}^{(1)}(a,b)=-\frac{1}{8\pi^{2}}\int_{0}^{\infty}ds\frac{e^{-m^{2}s}}{s^{3}}\left\{ (eas)\coth(eas)(ebs)\cot(ebs)-\left[1+\frac{(eas)^{2}-(ebs)^{2}}{3}\right]\right\} ,\label{L1ab_int}
\end{equation}
where $m$ is the electron mass, and $1+(es)^{2}\left(a^{2}-b^{2}\right)/3$
is subtracted to remove the zero-point energy and renormalize the
charge and fields for yielding a finite physical quantity \citep{Schwinger1951Gauge}.
This expression can be rewritten in a form convenient for the case
of $a\gg b$, i.e., $\tilde{b}=b/a\ll1$:

\begin{equation}
\mathcal{L}^{(1)}(a,b)=\mathcal{\bar{L}}^{(1)}(\bar{a},\tilde{b})=\frac{m^{4}}{8\pi^{2}}\frac{1}{4\bar{a}^{2}}\int_{0}^{\infty}\mathrm{d}z\frac{e^{-2\bar{a}z}}{z^{3}}\left[1+\frac{z^{2}(1-\tilde{b}^{2})}{3}-\tilde{b}z^{2}\coth(z)\cot(z\tilde{b})\right],\label{L1ab_int_z}
\end{equation}
where the dimensionless variable and parameters are

\begin{equation}
z=eas,\quad\bar{a}=\frac{m^{2}}{2ea},\quad\bar{b}=\frac{m^{2}}{2eb},\quad\tilde{b}=\frac{b}{a}=\frac{\bar{a}}{\bar{b}}.\label{abar_bbar_btilde}
\end{equation}
For a pure magnetic field, $\bar{a}=(B_{c}/B)/2$ and $\bar{b}=\infty$,
where $B_{c}=m^{2}/e=4.4\times10^{13}\,\mathrm{gauss}/\sqrt{4\pi}$
is the critical magnetic field strength.

\begin{figure}
\includegraphics[width=0.5\textwidth]{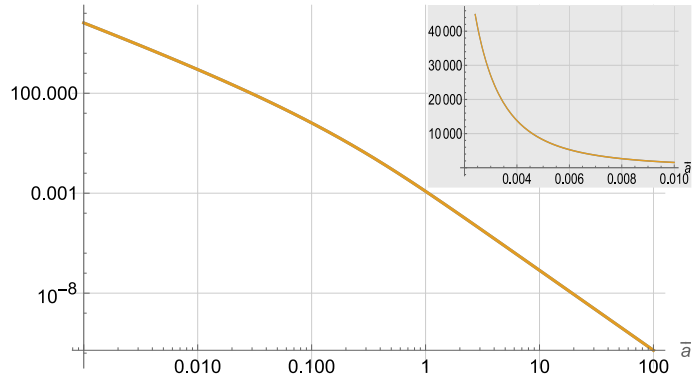}

\caption{Comparison of the integral and closed expressions of $\bar{\mathcal{L}}^{(1)}(\bar{a},0)$.
The integral expression is (\ref{L1ab_int_z}) with $\tilde{b}=0$,
and the closed one is (\ref{L1a_zeta}). The plotted values are in
units of $m^{4}/8\pi^{2}$. The parameter $\bar{a}$ ranges from 0.001
to 100, corresponding to $B/B_{c}$ from 500 to 0.005 for the purely
magnetic case. The inset shows the values in linear scale for the
range of $\bar{a}$ from 0.002 to 0.01. The plots of the two expressions
perfectly overlap each other. }

\label{fig:L1a_a_i}
\end{figure}

This integral can be numerically evaluated as described in App.~\ref{sec:Numerical},
but an explicit closed expression can be favored for theoretical analysis.
For the case of $b=0$, an explicit expression was obtained either
by the dimensional regularization of (\ref{L1ab_int}) \citep{Dittrich1976One}
or by the Schwinger-DeWitt in-out formalism combined with $\Gamma$-function
regularization \citep{Kim2019Quantum}:

\begin{equation}
\mathcal{\bar{L}}^{(1)}(\bar{a},0)\equiv\frac{m^{4}}{8\pi^{2}\bar{a}^{2}}H(\bar{a})=\frac{m^{4}}{8\pi^{2}\bar{a}^{2}}\left[\zeta'(-1,\bar{a})-\frac{1}{12}+\frac{\bar{a}^{2}}{4}-\left(\frac{1}{12}-\frac{\bar{a}}{2}+\frac{\bar{a}^{2}}{2}\right)\ln\bar{a}\right],\label{L1a_zeta}
\end{equation}
where $\zeta(s,\bar{a})$ is the Hurwitz zeta function, and $\zeta'(-1,\bar{a})=d\zeta(s,\bar{a})/ds|_{s=-1}$
\citep{Olver2010NIST} (See App.~\ref{app:La0} for more details.).
This expression perfectly matches the numerical evaluation of (\ref{L1ab_int_z})
with $\tilde{b}=0$, as shown in Fig.~\ref{fig:L1a_a_i}. We use
this analytic expression of $\mathcal{\bar{L}}^{(1)}(\bar{a},0)$
to express $\mathcal{\bar{L}}^{(1)}(\bar{a},\tilde{b})$ as a series
of $\tilde{b}=b/a$, as shown in the next section.

\subsection{Expansion of \textmd{\normalsize{}$\mathcal{\bar{L}}^{(1)}(\bar{a},\tilde{b})$}
in $\tilde{b}$\textmd{\normalsize{}\label{sec:Lab_exp}}}

In the magnetosphere of pulsars, the magnetic field
is much stronger than the electric field, and thus $a\gg b$, or equivalently
$\bar{a}\ll\bar{b}$, from (\ref{ab_FG}) and (\ref{FG_EB}). For
example, a pulsar model \citep{Goldreich1969Pulsar} gives $\tilde{b}$
($=b/a$) as a function decreasing with the distance from the pulsar
center: $\tilde{b}(R)\le0.2$ and $\tilde{b}(10R)\le0.02$, where
$R$ is the radius of the pulsar. For such a condition, the expansion
of $\mathcal{\bar{L}}^{(1)}(\bar{a},\tilde{b})$ in $\tilde{b}$ is
useful for analysis.

In the integral expression of $\mathcal{\bar{L}}^{(1)}(\bar{a},\tilde{b})$
(\ref{L1ab_int_z}), $\cot(\tilde{b}z)$ has poles at $z=n\pi/\tilde{b}$
($n=1,2,\dots$), which may apparently make the integral diverge.
However, only the Cauchy principal value of the integral is relevant
for vacuum birefringence, and the symmetric behavior of the cotangent
function around the poles prevents the principal value from diverging.
Furthermore, if $\exp(-2\bar{a}z)$ suppresses the integrand sufficiently
much before the first pole, i.e., $1/(2\bar{a})\ll\pi/\tilde{b}$,
or equivalently $\bar{b}\gg1/(2\pi)$, an asymptotic expression valid
for $\bar{b}\gg1/(2\pi)$ can be obtained. To proceed, we substitute
the series form of $(\tilde{b}z)\cdot\cot(\tilde{b}z)$ in (\ref{L1ab_int_z})
(4.19.6 of \citep{Olver2010NIST}):

\begin{equation}
(\tilde{b}z)\cdot\cot(\tilde{b}z)=\sum_{n=0}^{\infty}\frac{B_{2n}(-1)^{n}(2\tilde{b}z)^{2n}}{(2n)!},\label{ez_cotez}
\end{equation}
where $B_{2n}$ are the Bernoulli numbers. The series is convergent
only for $|\tilde{b}z|<\pi$ due to the nearest poles at $\tilde{b}z=\pm\pi$.
Upon substitution, the contribution outside the region of convergence
asymptotically vanishes as $\bar{b}\rightarrow\infty$. Then, the
integral in (\ref{L1ab_int_z}) is written as

\begin{equation}
\int_{0}^{\infty}\mathrm{d}z\frac{e^{-2\bar{a}z}}{z^{3}}\left\{ 1+\frac{z^{2}(1-\tilde{b}^{2})}{3}-z\coth(z)\sum_{n=0}^{\infty}\frac{(-1)^{n}B_{2n}(2\tilde{b}z)^{2n}}{(2n)!}\right\} .\label{L1ab_int_only}
\end{equation}

To evaluate the integral, we use the integral representation of $H(\bar{a})$
in (\ref{L1a_zeta}), which is derived by comparing (\ref{L1a_zeta})
with (\ref{L1a_int}):

\begin{equation}
H(\bar{a})=\frac{1}{4}\int_{0}^{\infty}\mathrm{d}z\frac{e^{-2\bar{a}z}}{z^{3}}\left\{ 1+\frac{z^{2}}{3}-z\coth(z)\right\} .\label{Ha_int}
\end{equation}
Differentiating (\ref{Ha_int}) $2n$ times by $\bar{a}$ yields a
useful formula:

\begin{equation}
\left(\frac{\mathrm{d}}{\mathrm{d}\bar{a}}\right)^{2n}H(\bar{a})\equiv H^{(2n)}(\bar{a})=\frac{2^{2n}}{4}\int_{0}^{\infty}\mathrm{d}z\frac{e^{-2\bar{a}z}}{z^{3}}\left\{ 1+\frac{z^{2}}{3}-z\coth(z)\right\} z^{2n},\label{H2na_int}
\end{equation}
of which closed form is given as (See App.~\ref{App:H2nz} for the
derivation)

\begin{equation}
H^{(2n)}(\bar{a})=\psi^{(2n-2)}(\bar{a})+\frac{1}{12}\frac{\Gamma(2n)}{\bar{a}^{2n}}+\frac{1}{2}\frac{\Gamma(2n-1)}{\bar{a}^{2n-1}}+\frac{\Gamma(2n-2)}{\bar{a}^{2n-2}}\theta(n-2)-\delta_{n1}\ln\bar{a},\quad(n\ge1),\label{H2na_anal}
\end{equation}
where $\psi^{(m)}(\bar{a})$ is the polygamma function. By using (\ref{H2na_int})
and (\ref{H2na_anal}) and the integral representation of the $\Gamma$
function, one can integrate (\ref{L1ab_int_only}) term-by-term to
obtain an asymptotic series of $\mathcal{\bar{\mathcal{L}}}^{(1)}(\bar{a},\tilde{b})$:

\begin{eqnarray}
\mathcal{\bar{\mathcal{L}}}^{(1)}(\bar{a},\tilde{b}) & = & \frac{m^{4}}{8\pi^{2}}\left[\frac{H(\bar{a})}{\bar{a}^{2}}+\left(\frac{1}{144\bar{a}^{4}}-\frac{H^{(2)}(\bar{a})}{12\bar{a}^{2}}\right)\tilde{b}^{2}\right]\nonumber \\
 &  & +\frac{m^{4}}{8\pi^{2}}\sum_{n=2}^{\infty}(-1)^{n}B_{2n}\left[\frac{H^{(2n)}(\bar{a})}{(2n)!\bar{a}^{2}}-\frac{1}{2n(2n-1)(2n-2)\bar{a}^{2n}}-\frac{1}{24n\bar{a}^{2n+2}}\right]\tilde{b}^{2n},\label{L1ab_series}
\end{eqnarray}
in which all the terms are given in terms of special functions. Thus,
(\ref{L1ab_series}) provides a systematic explicit expression of
$\mathcal{\bar{L}}^{(1)}(\bar{a},\tilde{b})$ in powers of $\tilde{b}=b/a$
for an arbitrary value of $\bar{a}$. For example, the first three
leading orders are given as

\begin{eqnarray}
\mathcal{\bar{\mathcal{L}}}^{(1)}(\bar{a},\tilde{b}) & = & \frac{m^{4}}{8\pi^{2}}\frac{1}{\bar{a}^{2}}\left[\zeta'(-1,\bar{a})-\frac{1}{12}+\frac{\bar{a}^{2}}{4}-\left(\frac{1}{12}-\frac{\bar{a}}{2}+\frac{\bar{a}^{2}}{2}\right)\ln\bar{a}\right]\label{L1ab_lowest3}\\
 &  & +\frac{m^{4}}{8\pi^{2}}\left(-\frac{1}{24\bar{a}^{3}}+\frac{\ln\bar{a}-\psi^{(0)}(\bar{a})}{12\bar{a}^{2}}\right)\text{\ensuremath{\tilde{b}^{2}}}+\frac{m^{4}}{8\pi^{2}}\left(-\frac{1}{720\bar{a}^{5}}-\frac{\psi^{(2)}(\bar{a})}{720\bar{a}^{2}}\right)\tilde{b}^{4}+O(\tilde{b}^{6}).\nonumber
\end{eqnarray}
A similar expansion in $K=-16a^{2}b^{2}$ was made by Heyl and Hernquist
to yield the explicit expressions of the lowest few orders \citep{Heyl1997Analytic}.

When both electric and magnetic fields are highly subcritical, i.e.,
$\bar{a}\gg1$ and $\bar{b}\gg1$ hold, the lowest order can be obtained
from the expansion up to $O(\tilde{b}^{4})$, and the second lowest
from the expansion up to $O(\tilde{b}^{6})$:

\begin{eqnarray}
\mathcal{\bar{\mathcal{L}}}^{(1)}(\bar{a},\tilde{b}) & = & \frac{m^{4}}{8\pi^{2}}\frac{1}{\bar{a}^{4}}\left[\frac{1}{720}+\frac{\tilde{b}^{2}}{144}+\frac{\tilde{b}^{4}}{720}\right]+\frac{m^{4}}{8\pi^{2}}\frac{1}{\bar{a}^{6}}\left[-\frac{1}{5040}-\frac{\tilde{b}^{2}}{1440}+\frac{\tilde{b}^{4}}{1440}+\frac{\tilde{b}^{6}}{5040}\right]\label{L1ab_weakO6}\\
 & = & \frac{m^{4}}{8\pi^{2}}\left[\frac{1}{720\bar{a}^{4}}+\frac{1}{144\bar{a}^{2}\bar{b}^{2}}+\frac{1}{720\bar{b}^{4}}\right]+\frac{m^{4}}{8\pi^{2}}\left[-\frac{1}{5040\bar{a}^{6}}-\frac{1}{1440\bar{a}^{4}\bar{b}^{2}}+\frac{1}{1440\bar{a}^{2}\bar{b}^{4}}+\frac{1}{5040\bar{b}^{6}}\right]\nonumber
\end{eqnarray}
The term with the first bracket is written in terms of $\mathbf{E}$
and $\mathbf{B}$ as

\begin{equation}
\frac{e^{4}}{360\pi^{2}m^{4}}\left[(\mathbf{B}^{2}-\mathbf{E}^{2})^{2}+7(\mathbf{E}\cdot\mathbf{B})^{2}\right],\label{L1ab_weakO4}
\end{equation}
which was obtained by Heisenberg and Euler \citep{Heisenberg1936Folgerungen},
and Schwinger \citep{Schwinger1951Gauge}.

It needs to be mentioned that the one-loop effective Lagrangian $\mathcal{L}^{(1)}(a,b)$
(\ref{L1ab_int}) can be expressed as a convergent series of which
terms are some special functions of $a$ and $b$ \citep{Cho2001Convergent,Jentschura2002QED,Cho2006Light}.
The expansion may be exact but is not very convenient for the condition
of an arbitrarily strong magnetic field combined with a weaker electric
field because the expansion needs an infinite sum. The convergence
of the sum is slow, and thus its evaluation needs acceleration techniques
\citep{Jentschura2002QED}. In this regards, the expansion (\ref{L1ab_lowest3})
is more convenient for theoretical analysis.

\subsection{Behavior of $\mathcal{\bar{\mathcal{L}}}^{(1)}(\bar{a},\tilde{b})$
and the validity of its expansion expression}

\begin{figure}
\includegraphics[width=0.5\textwidth]{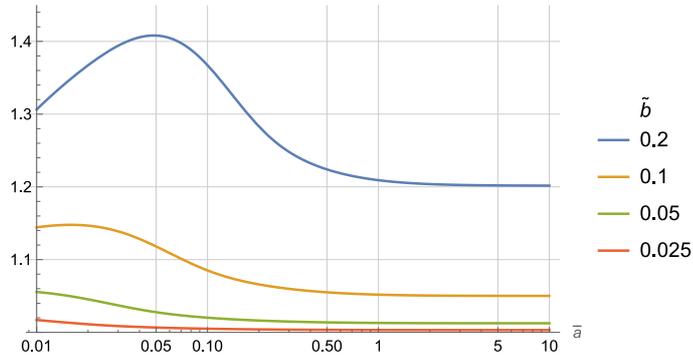}

\caption{$\mathcal{\bar{\mathcal{L}}}^{(1)}(\bar{a},\tilde{b})/\mathcal{\bar{\mathcal{L}}}^{(1)}(\bar{a},0)$
for $\tilde{b}=0.2,$0.1, 0.05, 0.025. $\mathcal{\bar{\mathcal{L}}}^{(1)}(\bar{a},\tilde{b})$
was obtained by the direct integration of (\ref{L1ab_int_z}), in
which the first 10 poles were included.}

\label{fig:L1int_bt_varied}
\end{figure}

The exact dependence of $\mathcal{\bar{\mathcal{L}}}^{(1)}(\bar{a},\tilde{b})$
with $\tilde{b}$ can be investigated by numerically evaluating the
integral expression (\ref{L1ab_int_z}), as described in App.~\ref{sec:Numerical}.
As $\mathcal{\bar{\mathcal{L}}}^{(1)}(\bar{a},0)$ is completely known,
the ratio $\mathcal{\bar{\mathcal{L}}}^{(1)}(\bar{a},\tilde{b})/\mathcal{\bar{\mathcal{L}}}^{(1)}(\bar{a},0)$
represents the dependence on $\tilde{b}$ alone. In Fig.~\ref{fig:L1int_bt_varied},
the offset difference of $\mathcal{\bar{\mathcal{L}}}^{(1)}(\bar{a},\tilde{b})$
from $\mathcal{\bar{\mathcal{L}}}^{(1)}(\bar{a},0)$ increases with
$\tilde{b}$. Furthermore, additional significant differences appear
as $\bar{a}$ decreases below certain onset values. For example, when
$\tilde{b}=0.2$, the ratio is close 1.2 until $\bar{a}$ decreases
to 1, but, as $\bar{a}$ decreases below 1, it significantly increases
and then decreases . When $\tilde{b}$ is smaller, the behavior is
similar except that the offset difference is smaller, and the onset
value of $\bar{a}$ decreases. This additional difference is considered
to be attributed to the pair production. When $1/(2\bar{a})\gtrsim\pi/\tilde{b}$,
or equivalently $\bar{b}\apprle1/(2\pi)$, the contribution from the
first pole of the cotangent function in (\ref{L1ab_int_z}) becomes
significant. As $\bar{b}=E_{\mathrm{cr}}/(2E)$ in the parallel field
configuration, the condition $\bar{b}\apprle1/(2\pi)$ implies the
emergence of the pair production by the strong electric field. In
such a case, the effective Lagrangian develops an imaginary part,
which is accompanied by a variation in the real part, i.e., the additional
difference in Fig.~\ref{fig:L1int_bt_varied}. It is similar to the
anomalous dispersion near a resonance region in the linear optical
response \citep{Born1999Principles}. As the imaginary part gives
the pair production probability, the plasma effects appears when the
probability is sufficiently high. Then, it is no more pure vacuum
birefringence. In our analysis, we focus on real part of the Lagrangian
to study the pure vacuum birefringence.

\begin{figure}
\includegraphics[width=1\textwidth]{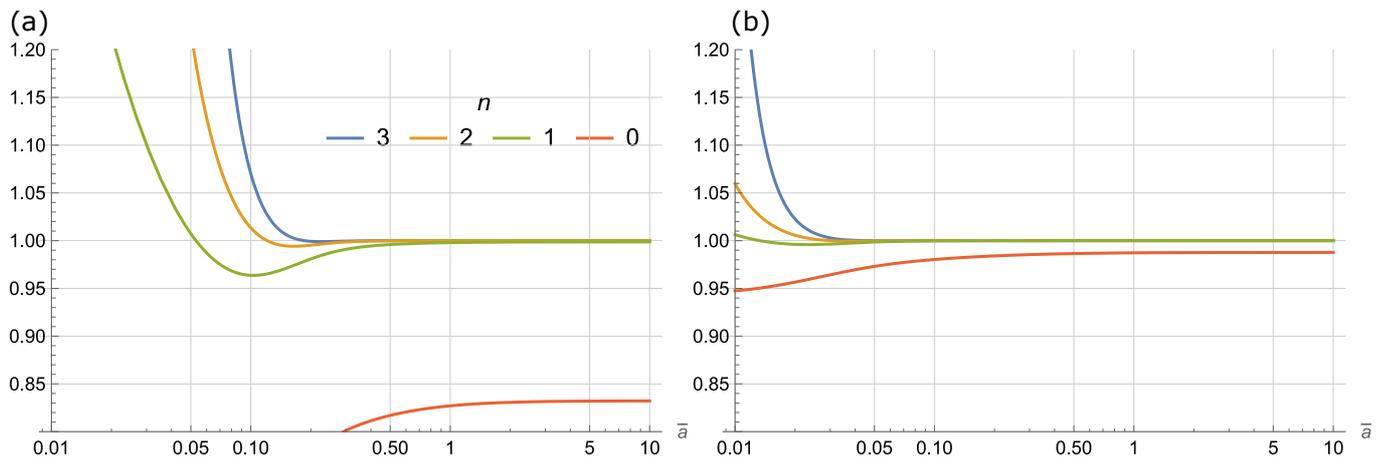}

\caption{$\mathcal{\bar{\mathcal{L}}}_{\mathrm{exp}}^{(1)}(\bar{a},\tilde{b},n)/\mathcal{\bar{\mathcal{L}}}_{\mathrm{int}}^{(1)}(\bar{a},\tilde{b})$
for $n=0,1,2,3$ when (a) $\tilde{b}=0.2$ and (b) $\tilde{b}=0.05$.
$\mathcal{\bar{\mathcal{L}}}_{\mathrm{exp}}^{(1)}(\bar{a},\tilde{b},n)$
is the expansion (\ref{L1ab_series}) up to $O(\tilde{b}^{2n})$,
and $\mathcal{\bar{\mathcal{L}}}_{\mathrm{int}}^{(1)}(\bar{a},\tilde{b})$
is the numerical evaluation of (\ref{L1ab_int_z}), including the
first 10 poles.}

\label{fig:L1exp_n_varied}
\end{figure}

By using the numerical evaluation of $\mathcal{\bar{\mathcal{L}}}^{(1)}(\bar{a},\tilde{b})$,
we can find the parameter range in which the expansion (\ref{L1ab_series})
is accurate. The ratio $\mathcal{\bar{\mathcal{L}}}_{\mathrm{exp}}^{(1)}(\bar{a},\tilde{b},n)/\mathcal{\bar{\mathcal{L}}}_{\mathrm{int}}^{(1)}(\bar{a},\tilde{b})$
is plotted for $n=0,1,2,3$ and $\tilde{b}=0.2,0.05$, where $\mathcal{\bar{\mathcal{L}}}_{\mathrm{exp}}^{(1)}(\bar{a},\tilde{b},n)$
is the expansion (\ref{L1ab_series}) up to $O(\tilde{b}^{2n})$,
and $\mathcal{\bar{\mathcal{L}}}_{\mathrm{int}}^{(1)}(\bar{a},\tilde{b})$
is the numerical evaluation of (\ref{L1ab_int_z}). In Fig.~\ref{fig:L1exp_n_varied}(a),
$\mathcal{\bar{\mathcal{L}}}_{\mathrm{exp}}^{(1)}(\bar{a},\tilde{b}=0.2,n\ge1)$
is accurate within 1\% for $\bar{a}\ge0.3$. The 1\%-accuracy threshold
of $\bar{a}$, denoted by $\bar{a}_{1\%}$, decreases slightly as
$n$ increases, but a higher-$n$ expansion blows up faster below
$\bar{a}_{1\%}$; a typical behavior of the Taylor expansion. The
$n=0$ expansion, corresponding to the effective Lagrangian with $\tilde{b}=0$,
has an error larger than 17\%. At a lower value of $\tilde{b}=0.05$,
the overall behavior is similar to the case of $\tilde{b}=0.2$, but
$\bar{a}_{1\%}$ is lowered to 0.025, and the minimum error of the
$n=0$ expansion decreases to 2\%, as shown in Fig.~\ref{fig:L1exp_n_varied}(b).
In practice, the $n=2$ expansion can be a reasonable choice for analysis,
showing $\bar{a}_{1\%}=0.1$ for $\tilde{b}=0.2$ ( $B\le5B_{\mathrm{cr}}$
and $E\le E_{\mathrm{cr}}$) and $\bar{a}_{1\%}=0.02$ for $\tilde{b}=0.05$
( $B\le25B_{\mathrm{cr}}$ and $E\le1.25E_{\mathrm{cr}}$). Noting
that the expansion is not accurate when the pair production is significant,
one may roughly estimate a threshold value of $\bar{a}$ like $\bar{a}_{1\%}$
by requiring $\tilde{b}/(2\bar{a}_{\mathrm{thres}})\simeq1$. As $\tilde{b}/(2\bar{a})=E/E_{\mathrm{cr}}$
in the parallel field configuration, the expansion would not be accurate
when this ratio is the order of unity. In Fig.~\ref{fig:L1exp_n_varied},
the threshold values $\bar{a}_{1\%}$ are close to the estimation
$\bar{a}_{\mathrm{thres}}=\tilde{b}/2$.

\section{Response of the vacuum in strong electromagnetic fields\label{sec:Response}}

In contrast to the classical vacuum, the quantum vacuum in electromagnetic
fields behaves as a medium, of which response is quantified by the
polarization ($\mathbf{P}$) and the magnetization ($\mathbf{M}$),
or equivalently by the electric induction ($\mathbf{D}$) and magnetic
field strength ($\mathbf{H}$). These quantities can be obtained by
considering the variation of the effective Lagrangian with respect
to that of the electromagnetic field \citep{Heisenberg1936Folgerungen,Berestetskii1982Quantum}:

\begin{equation}
\mathbf{D}=\mathbf{E}+\mathbf{P}=\frac{\partial\mathcal{L}_{\mathrm{eff}}}{\partial\mathbf{E}}=\mathbf{E}+\frac{\partial\mathcal{L}^{(1)}}{\partial\mathbf{E}},\quad\mathbf{H}=\mathbf{B}-\mathbf{M}=-\frac{\partial\mathcal{L}_{\mathrm{eff}}}{\partial\mathbf{B}}=\mathbf{B}-\frac{\partial\mathcal{L}^{(1)}}{\partial\mathbf{B}}.\label{DHPM_def}
\end{equation}
In this section, we derive expressions of $\mathbf{P}$ and $\mathbf{M}$
for an arbitrary $\mathcal{L}^{(1)}(a,b)$ and use the result to obtain
the permittivity and permeability tensors for weak low-frequency ($\omega\ll m$)
probe fields. These tensors are necessary to analyze the vacuum birefringence
in the next section. These results can be obtained also in a explicitly
Lorentz covariant manner by using photon polarization tensors \citep{BialynickaBirula1970Nonlinear,Dittrich2000Probing,Melrose2013Quantum,Karbstein2015Photon}.

\subsection{Polarization and magnetization of the vacuum in uniform electric
and magnetic fields}

The effective Lagrangian is a function of $a$ and $b$ (\ref{ab_FG})
and (\ref{FG_EB}), and thus we consider the variation of a general
differentiable function $f(a,b)$:

\begin{equation}
\delta f=\delta a\cdot\partial_{a}f+\delta b\cdot\partial_{b}f.
\end{equation}
As $a$ and $b$ are functions of $F$ and $G$, the variations $\delta a$
and $\delta b$ can be written in terms of the variations $\delta F$
and $\delta G$:

\begin{equation}
\delta a=\frac{\partial a}{\partial F}\delta F+\frac{\partial a}{\partial G}\delta G,\quad\delta b=\frac{\partial b}{\partial F}\delta F+\frac{\partial b}{\partial G}\delta G.\label{da_db}
\end{equation}
In turn, the variations of $\delta F$ and $\delta G$ can also be
written in terms of $\delta\mathbf{E}$ and $\delta\mathbf{B}$:

\begin{equation}
\delta F=\frac{\partial F}{\partial\mathbf{E}}\cdot\delta\mathbf{E}+\frac{\partial F}{\partial\mathbf{B}}\cdot\delta\mathbf{B},\quad\delta G=\frac{\partial G}{\partial\mathbf{E}}\cdot\delta\mathbf{E}+\frac{\partial G}{\partial\mathbf{B}}\cdot\delta\mathbf{B}.\label{dF_dG}
\end{equation}
Thus, usign the chain rules, the coefficients in these variations
can be calculated by using (\ref{ab_FG}) and (\ref{FG_EB}) to express
$\delta f$ in terms of $\delta\mathbf{E}$ and $\delta\mathbf{B}$:

\begin{equation}
\delta f=\delta\mathbf{E}\cdot\left(-\mathbf{E}\hat{S}-\mathbf{B}\hat{A}\right)f+\delta\mathbf{B}\cdot\left(\mathbf{B}\hat{S}-\mathbf{E}\hat{A}\right)f,\label{df}
\end{equation}
where

\begin{equation}
\hat{S}=\frac{\left(a\cdot\partial_{a}-b\cdot\partial_{b}\right)}{a^{2}+b^{2}}=\partial_{F},\quad\hat{A}=\frac{\sigma\left(b\cdot\partial_{a}+a\cdot\partial_{b}\right)}{a^{2}+b^{2}}=\partial_{G},\quad\hat{S}\hat{A}=\hat{A}\hat{S}=\partial_{F}\partial_{G}.\label{SA_ab}
\end{equation}
The operators $\hat{S}$ and $\hat{A}$ are symmetric and antisymmetric
under the parity inversion, respectively. Note that $(a^{2}+b^{2})\hat{S}$
measures the difference of homogeneity of polynomials such that $(a^{2}+b^{2})\hat{S}(a^{m}b^{n})=(m-n)(a^{m}b^{n})$,
while $(a^{2}+b^{2})\hat{A}$ measures the mixed homogeneity of polynomials
such that $(a^{2}+b^{2})\hat{A}(a^{m}b^{n})=(mb^{2}+na^{2})(a^{m-1}b^{n-1})$.
In addition, $(a^{2}+b^{2})\hat{S}$ preserves the polynomial order
$(m,n)$, while $(a^{2}+b^{2})\hat{A}$ changes it to $(m-1,n+1)$
and $(m+1,n-1)$ while maintaining the sum of the orders of $a$ and
$b$. Replacing $f$ with $\mathcal{L}^{(1)}$ and using (\ref{DHPM_def}),
one can obtain $\mathbf{P}$ and $\mathbf{M}$ from $\mathcal{L}^{(1)}(a,b)$:

\begin{equation}
\mathbf{P}=\frac{\partial\mathcal{L}^{(1)}}{\partial\mathbf{E}}=-\mathbf{E}\hat{S}\mathcal{L}^{(1)}-\mathbf{B}\hat{A}\mathcal{L}^{(1)},\quad\mathbf{M}=\frac{\partial\mathcal{L}^{(1)}}{\partial\mathbf{B}}=\mathbf{B}\hat{S}\mathcal{L}^{(1)}-\mathbf{E}\hat{A}\mathcal{L}^{(1)}.\label{PM_L1ab}
\end{equation}
Despite the appearance, this relation is not linear in $\mathbf{E}$
and $\mathbf{B}$, as $\hat{S}\mathcal{L}^{(1)}$ and $\hat{A}\mathcal{L}^{(1)}$
are nonlinear functions of $\mathbf{E}$ and $\mathbf{B}$.

\subsection{Permittivity and permeability tensors for weak low-frequency probe
fields}

The material-like vacuum with the nonlinear response relation (\ref{PM_L1ab})
affects the propagation of photons. When the photon energy is much
smaller than the electron's rest mass energy ($\omega\ll m=0.5\,\mathrm{MeV}$),
the photon can be treated as a weak perturbation field (probe field)
added to the strong background field \citep{Mckenna1963Nonlinear,Adler1971Photon}.
Then, $\mathbf{E}$, $\mathbf{B}$, $\mathbf{D}$, $\mathbf{H}$,
$\mathbf{P}$, and $\mathbf{M}$ can be decomposed as

\begin{equation}
\mathbf{A}=\mathbf{A}_{0}+\delta\mathbf{A},
\end{equation}
where $\mathbf{A}_{0}$ ($\delta\mathbf{A}$) refers to the quantities
of the background (probe) field. Furthermore, as $\mathbf{A}_{0}$
is uniform, while $\delta\mathbf{A}$ is varying, the relation (\ref{DHPM_def})
should be satisfied separately between the uniform and varying quantities:

\begin{equation}
\mathbf{D}_{0}=\mathbf{E}_{0}+\mathbf{P}_{0},\quad\mathbf{H}_{0}=\mathbf{B}_{0}-\mathbf{M}_{0},
\end{equation}

\begin{equation}
\delta\mathbf{D}=\delta\mathbf{E}+\delta\mathbf{P},\quad\delta\mathbf{H}=\delta\mathbf{B}-\delta\mathbf{M}.
\end{equation}
Then, $\delta\mathbf{D}$ and $\delta\mathbf{H}$ can be obtained
by varying the relation (\ref{PM_L1ab}) around $(\mathbf{E}_{0},\mathbf{B}_{0})$,
for which the formula (\ref{df}) is convenient. A straightforward
but lengthy calculation yields the linear relations among the varying
quantities relevant to the probe field:

\begin{equation}
\delta\mathbf{D}=\boldsymbol{\epsilon}_{E}\cdot\delta\mathbf{E}+\boldsymbol{\epsilon}_{B}\cdot\delta\mathbf{B},\quad\delta\mathbf{H}=\bar{\boldsymbol{\mu}}_{B}\cdot\delta\mathbf{B}+\bar{\boldsymbol{\mu}}_{E}\cdot\delta\mathbf{E},\label{dDdH}
\end{equation}
where $\boldsymbol{\epsilon}_{E}$, $\boldsymbol{\epsilon}_{B}$,
$\bar{\boldsymbol{\mu}}_{B}$, and $\bar{\boldsymbol{\mu}}_{E}$ are
3-by-3 tensors. Their components are as follows:

\begin{equation}
\epsilon_{E,ij}=\delta_{ij}(1-\mathcal{L}_{S}^{(1)})+\left[E_{0i}E_{0j}\mathcal{L}_{SS}^{(1)}+\left(E_{0i}B_{0j}+B_{0i}E_{0j}\right)\mathcal{L}_{SA}^{(1)}+B_{0i}B_{0j}\mathcal{L}_{AA}^{(1)}\right],\label{eEij}
\end{equation}

\begin{equation}
\epsilon_{B,ij}=-\delta_{ij}\mathcal{L}_{S}^{(1)}+\left[-E_{0i}B_{0j}\mathcal{L}_{SS}^{(1)}+\left(E_{0i}E_{0j}-B_{0i}B_{0j}\right)\mathcal{L}_{SA}^{(1)}+B_{0i}E_{0j}\mathcal{L}_{AA}^{(1)}\right],\label{eBij}
\end{equation}

\begin{equation}
\bar{\mu}_{B,ij}=\delta_{ij}(1-\mathcal{L}_{S}^{(1)})+\left[-B_{0i}B_{0j}\mathcal{L}_{SS}^{(1)}+\left(B_{0i}E_{0j}+E_{0i}B_{0j}\right)\mathcal{L}_{SA}^{(1)}-E_{0i}E_{0j}\mathcal{L}_{AA}^{(1)}\right],\label{muBij}
\end{equation}

\begin{equation}
\bar{\mu}_{E,ij}=\delta_{ij}\mathcal{L}_{A}^{(1)}+\left[B_{0i}E_{0j}\mathcal{L}_{SS}^{(1)}+\left(B_{0i}B_{0j}-E_{0i}E_{0j}\right)\mathcal{L}_{SA}^{(1)}-E_{0i}B_{0j}\mathcal{L}_{AA}^{(1)}\right],\label{muEij}
\end{equation}
where

\begin{equation}
\mathcal{L}_{S}^{(1)}=\hat{S}_{0}\mathcal{L}^{(1)},\quad\mathcal{L}_{A}^{(1)}=\hat{A}_{0}\mathcal{L}^{(1)},\quad\mathcal{L}_{SS}^{(1)}=\left(\hat{S}\hat{S}\right)_{0}\mathcal{L}^{(1)},\quad\mathcal{L}_{SA}^{(1)}=\left(\hat{S}\hat{A}\right)_{0}\mathcal{L}^{(1)},\quad\mathcal{L}_{AA}^{(1)}=\left(\hat{A}\hat{A}\right)_{0}\mathcal{L}^{(1)}.\label{SA_L1}
\end{equation}
Here, the subscript 0 means that $\hat{S}$ and $\hat{A}$ are evaluated
at $(\mathbf{E}_{0},\mathbf{B}_{0})$. Note that these formulas are
valid for any effective Lagrangian. From now on, we shall focus on
the Heisenberg-Euler and Schwinger effective Lagrangian (\ref{L1ab_int}).

The tensors $\boldsymbol{\epsilon}_{B}$ and $\bar{\boldsymbol{\mu}}_{E}$
represent the magneto-electric response to the probe field, in which
a magnetic (electric) field induces polarization (magnetization) \citep{Melrose1991Electromagnetic},
unlike usual dielectric and magnetic materials (See \citep{Fiebig2005Revival,Eerenstein2006Multiferroic}
for the recent studies of the magneto-electric response in condensed
matter). Such response disappears as the background electric field
vanishes: when $\mathbf{E}_{0}=0$, $\mathcal{L}_{A}^{(1)}$ and $\mathcal{L}_{SA}^{(1)}$
in (\ref{eBij},\ref{muEij}) vanishes for the Lagrangian (\ref{L1ab_int})
to yield $\boldsymbol{\epsilon}_{B}=\bar{\boldsymbol{\mu}}_{E}=0$.

The quantities in (\ref{SA_L1}) have the complete information for
the analysis of vacuum birefringence because they immediately lead
to the permittivity and permeability for an arbitrary background field
configuration through (\ref{eEij},\ref{eBij},\ref{muBij},\ref{muEij}).
By using the expansion (\ref{L1ab_lowest3}), we can obtain the formulas
of (\ref{SA_L1}) contributed from each order of $\tilde{b}$. For
example, the contributions from the lowest three orders are shown
in Tabs.~\ref{tab:SAL1_n0}, \ref{tab:SAL1_n1}, and \ref{tab:SAL1_n2}.
In Tab.~\ref{tab:SAL1_n0}, the dependence on $\bar{b}$ appears
due to the $b\cdot\partial_{a}$ term in $\hat{A}$ although $\mathcal{\bar{\mathcal{L}}}^{(1)}(\bar{a},0)$
does not depend on $\bar{b}$. As the combined field of the background
field and the probe field is not wrenchless in general, the expression
of $\mathcal{\bar{\mathcal{L}}}^{(1)}(\bar{a},\tilde{b}\neq0)$ is
necessary for wrenchless background fields.

For the case of $b=0$ with an arbitrary value of $a$, i.e., the
wrenchless case, the exact expressions of $\mathcal{L}_{S}^{(1)}$,
$\mathcal{L}_{A}^{(1)}$, $\mathcal{L}_{SS}^{(1)}$, $\mathcal{L}_{SA}^{(1)}$,
and $\mathcal{L}_{AA}^{(1)}$ can be obtained first by combining the
results in Tabs.~\ref{tab:SAL1_n0} and \ref{tab:SAL1_n1} and then
by taking the limit of $\bar{b}=\infty$. Higher order contributions
are absent in this case. The results are shown in Tab.~\ref{tab:SAL1_n1binf}.
The same results were obtained by Karbstein et al.~\citep{Karbstein2015Photon},
who expanded the one-loop effective action up to the second order
of the probe field and specified the calculation to the wrenchless
case.

\begin{table}
\begin{tabular}{|c|c|l|}
\hline
$\mathcal{L}_{S}^{(1)}$ & $-\frac{\bar{b}^{2}e^{2}}{24\left(\text{\ensuremath{\bar{a}}}^{2}+\bar{b}^{2}\right)\pi^{2}}$ & $6\text{\ensuremath{\bar{a}}}^{2}-6\ln(2\pi)\bar{a}+12\ln(\Gamma(\bar{a}))\bar{a}+(2-6\bar{a})\ln(\bar{a})-24\zeta'(-1,\bar{a})+1$\tabularnewline
\hline
$\mathcal{L}_{A}^{(1)}$ & $-\frac{\bar{a}\bar{b}e^{2}\sigma}{24\left(\text{\ensuremath{\bar{a}}}^{2}+\bar{b}^{2}\right)\pi^{2}}$ & $6\text{\ensuremath{\bar{a}}}^{2}-6\ln(2\pi)\text{\ensuremath{\bar{a}}}+12\ln(\Gamma(\bar{a}))\bar{a}+(2-6\bar{a})\ln(\bar{a})-24\zeta'(-1,\bar{a})+1$\tabularnewline
\hline
\multirow{2}{*}{$\mathcal{L}_{SS}^{(1)}$} & \multirow{2}{*}{$\frac{\text{\ensuremath{\bar{a}}}^{2}\bar{b}^{4}e^{4}}{3\left(\bar{a}^{2}+\bar{b}^{2}\right)^{3}m^{4}\pi^{2}}$} & $6(\psi^{(0)}(\bar{a})+1)\text{\ensuremath{\bar{a}}}^{4}+6\bar{a}^{4}+9\ln(\bar{a})\text{\ensuremath{\bar{a}}}^{3}-12\text{\ensuremath{\bar{a}}}^{3}-6\bar{b}^{2}\text{\ensuremath{\bar{a}}}^{2}-4\ln(\text{\ensuremath{\bar{a}}})\bar{a}^{2}+6\bar{b}^{2}(\psi^{(0)}(\bar{a})+1)\bar{a}^{2}$\tabularnewline
 &  & $+48\zeta'(-1,\bar{a})\bar{a}^{2}-\bar{a}^{2}-3\bar{b}^{2}\ln(\bar{a})\bar{a}-3\left(5\text{\ensuremath{\bar{a}}}^{2}+\bar{b}^{2}\right)(2\bar{a}-\ln(2\pi)+2\ln(\Gamma(\bar{a}))-1)\bar{a}+\bar{b}^{2}$\tabularnewline
\hline
\multirow{2}{*}{$\mathcal{L}_{SA}^{(1)}$} & \multirow{2}{*}{$\frac{\bar{a}^{3}\bar{b}^{3}e^{4}\sigma}{3\left(\bar{a}^{2}+\bar{b}^{2}\right)^{3}m^{4}\pi^{2}}$} & $-12\bar{a}^{4}+3\ln(\bar{a})\bar{a}^{3}+9\ln(2\pi)\text{\ensuremath{\bar{a}}}^{3}-18\ln(\Gamma(\text{\ensuremath{\bar{a}}}))\bar{a}^{3}+3\bar{a}^{3}-2\ln(\bar{a})\bar{a}^{2}+6\left(\bar{a}^{2}+\bar{b}^{2}\right)\psi^{(0)}(\bar{a})\bar{a}^{2}$\tabularnewline
 &  & $+3\bar{b}^{2}\text{\ensuremath{\bar{a}}}-9\bar{b}^{2}\ln(\bar{a})\bar{a}-3\bar{b}^{2}\ln(2\pi)\bar{a}+6\bar{b}^{2}\ln(\Gamma(\bar{a}))\bar{a}+2\bar{b}^{2}+2\bar{b}^{2}\ln(\bar{a})+24\left(\bar{a}^{2}-\bar{b}^{2}\right)\zeta'(-1,\bar{a})$\tabularnewline
\hline
\multirow{3}{*}{$\mathcal{L}_{AA}^{(1)}$} & \multirow{3}{*}{$\frac{\bar{a}^{2}\bar{b}^{2}e^{4}\sigma^{2}}{6\left(\bar{a}^{2}+\bar{b}^{2}\right)^{3}m^{4}\pi^{2}}$} & $-18\bar{a}^{6}+12\ln(2\pi)\bar{a}^{5}-24\ln(\Gamma(\bar{a}))\bar{a}^{5}+6\bar{a}^{5}-2\ln(\bar{a})\bar{a}^{4}+12\left(\bar{a}^{2}+\bar{b}^{2}\right)\psi^{(0)}(\bar{a})\bar{a}^{4}+\bar{a}^{4}$\tabularnewline
 &  & $+6\bar{b}^{2}\bar{a}^{3}-18\bar{b}^{2}\ln(\bar{a})\bar{a}^{3}-6\bar{b}^{2}\ln(2\pi)\bar{a}^{3}+12\bar{b}^{2}\ln(\Gamma(\bar{a}))\bar{a}^{3}-6\bar{b}^{4}\bar{a}^{2}+4\bar{b}^{2}\bar{a}^{2}+4\bar{b}^{2}\ln(\bar{a})\bar{a}^{2}+6\bar{b}^{4}\ln(\bar{a})\text{\ensuremath{\bar{a}}}$\tabularnewline
 &  & $+6\bar{b}^{4}\ln(2\pi)\text{\ensuremath{\bar{a}}}-12\bar{b}^{4}\ln(\Gamma(\bar{a}))\bar{a}-\bar{b}^{4}-2\bar{b}^{4}\ln(\text{\ensuremath{\bar{a}}})+24\left(\bar{a}^{2}-\bar{b}^{2}\right)^{2}\zeta'(-1,\bar{a})$\tabularnewline
\hline
\end{tabular}

\caption{Contribution to $\mathcal{L}_{S}^{(1)}$, $\mathcal{L}_{A}^{(1)}$,
$\mathcal{L}_{SS}^{(1)}$, $\mathcal{L}_{SA}^{(1)}$, and $\mathcal{L}_{AA}^{(1)}$
from the $O(\tilde{b}^{0})$ term in the expansion of $\mathcal{\bar{\mathcal{L}}}^{(1)}(\bar{a},\tilde{b})$
(\ref{L1ab_lowest3}). Each factor in the second column should be
multiplied to the corresponding formula in the third column. $\bar{b}=\bar{a}/\tilde{b}$. }

\label{tab:SAL1_n0}
\end{table}

\begin{table}
\begin{tabular}{|c|c|l|}
\hline
$\mathcal{L}_{S}^{(1)}$ & $\frac{\bar{a}e^{2}}{48\left(\bar{a}^{2}+\bar{b}^{2}\right)\pi^{2}}$ & $2\psi^{(1)}(\bar{a})\bar{a}^{2}-4\ln(\bar{a})\bar{a}+4\psi^{(0)}(\bar{a})\bar{a}-2\bar{a}+1$\tabularnewline
\hline
$\mathcal{L}_{A}^{(1)}$ & $\frac{e^{2}\sigma}{48\tilde{b}\left(\bar{a}^{2}+\bar{b}^{2}\right)\pi^{2}}$ & $2\psi^{(1)}(\bar{a})\bar{a}^{4}-2\bar{a}^{3}-\bar{a}^{2}+4\bar{b}^{2}\ln(\bar{a})\bar{a}-4\bar{b}^{2}\psi^{(0)}(\bar{a})\bar{a}-2\bar{b}^{2}$\tabularnewline
\hline
\multirow{2}{*}{$\mathcal{L}_{SS}^{(1)}$} & \multirow{2}{*}{$-\frac{\bar{a}^{3}\bar{b}^{2}e^{4}}{12\left(\bar{a}^{2}+\bar{b}^{2}\right)^{3}m^{4}\pi^{2}}$} & $2\psi^{(2)}(\bar{a})\bar{a}^{5}+2\bar{b}^{2}\psi^{(2)}(\bar{a})\bar{a}^{3}-4\text{\ensuremath{\bar{a}}}^{3}-\bar{a}^{2}-12\bar{b}^{2}\bar{a}-16\bar{b}^{2}\ln(\bar{a})\bar{a}+16\bar{b}^{2}\psi^{(0)}(\bar{a})\bar{a}+3\bar{b}^{2}$\tabularnewline
 &  & $+2\left(3\bar{a}^{4}+7\bar{b}^{2}\bar{a}^{2}\right)\psi^{(1)}(\bar{a})$\tabularnewline
\hline
\multirow{2}{*}{$\mathcal{L}_{SA}^{(1)}$} & \multirow{2}{*}{$-\frac{\bar{a}^{2}\bar{b}e^{4}\sigma}{12\left(\bar{a}^{2}+\bar{b}^{2}\right)^{3}m^{4}\pi^{2}}$} & $2\psi^{(2)}(\bar{a})\bar{a}^{7}+2\bar{b}^{2}\psi^{(2)}(\bar{a})\bar{a}^{5}-4\bar{a}^{5}-\bar{a}^{4}-8\bar{b}^{2}\bar{a}^{3}-8\bar{b}^{2}\ln(\bar{a})\bar{a}^{3}+\bar{b}^{2}\text{\ensuremath{\bar{a}}}^{2}+4\bar{b}^{4}\bar{a}$\tabularnewline
 &  & $+8\bar{b}^{4}\ln(\bar{a})\bar{a}+8\bar{b}^{2}\left(\text{\ensuremath{\bar{a}}}^{2}-\bar{b}^{2}\right)\psi^{(0)}(\bar{a})\bar{a}-2\bar{b}^{4}+2\left(3\bar{a}^{6}+5\bar{b}^{2}\bar{a}^{4}-2\bar{b}^{4}\bar{a}^{2}\right)\psi^{(1)}(\bar{a})$\tabularnewline
\hline
\multirow{2}{*}{$\mathcal{L}_{AA}^{(1)}$} & \multirow{2}{*}{$-\frac{\bar{a}e^{4}\sigma^{2}}{12\left(\bar{a}^{2}+\bar{b}^{2}\right)^{3}m^{4}\pi^{2}}$} & $2\psi^{(2)}(\bar{a})\bar{a}^{9}+2\bar{b}^{2}\psi^{(2)}(\bar{a})\bar{a}^{7}-2\bar{a}^{7}-4\bar{b}^{2}\ln(\bar{a})\bar{a}^{5}+3\bar{b}^{2}\bar{a}^{4}+10\bar{b}^{4}\bar{a}^{3}+8\bar{b}^{4}\ln(\bar{a})\bar{a}^{3}$\tabularnewline
 &  & $+\bar{b}^{4}\bar{a}^{2}-4\bar{b}^{6}\ln(\bar{a})\bar{a}+4\bar{b}^{2}\left(\bar{a}^{2}-\bar{b}^{2}\right)^{2}\psi^{(0)}(\bar{a})\bar{a}+2\bar{b}^{6}+2\left(2\bar{a}^{8}+\bar{b}^{2}\bar{a}^{6}-5\bar{b}^{4}\bar{a}^{4}\right)\psi^{(1)}(\bar{a})$\tabularnewline
\hline
\end{tabular}

\caption{Contribution to $\mathcal{L}_{S}^{(1)}$, $\mathcal{L}_{A}^{(1)}$,
$\mathcal{L}_{SS}^{(1)}$, $\mathcal{L}_{SA}^{(1)}$, and $\mathcal{L}_{AA}^{(1)}$
from the $O(\tilde{b}^{2})$ term in the expansion of $\mathcal{\bar{\mathcal{L}}}^{(1)}(\bar{a},\tilde{b})$
(\ref{L1ab_lowest3}). Each factor in the second column should be
multiplied to the corresponding formula in the third column. $\bar{b}=\bar{a}/\tilde{b}$. }

\label{tab:SAL1_n1}
\end{table}

\begin{table}
\begin{tabular}{|c|c|l|}
\hline
$\mathcal{L}_{S}^{(1)}$ & $\frac{\bar{a}e^{2}}{1440\bar{b}^{2}\left(\bar{a}^{2}+\bar{b}^{2}\right)\pi^{2}}$ & $\psi^{(3)}(\bar{a})\bar{a}^{4}+6\psi^{(2)}(\bar{a})\bar{a}^{3}+3$\tabularnewline
\hline
$\mathcal{L}_{A}^{(1)}$ & $\frac{e^{2}\sigma}{1440\bar{b}^{3}\left(\bar{a}^{2}+\bar{b}^{2}\right)\pi^{2}}$ & $\psi^{(3)}(\bar{a})\bar{a}^{6}-\bar{a}^{2}-4\bar{b}^{2}+2\left(\bar{a}^{5}-2\bar{a}^{3}\bar{b}^{2}\right)\psi^{(2)}(\bar{a})$\tabularnewline
\hline
$\mathcal{L}_{SS}^{(1)}$ & $-\frac{\bar{a}^{3}e^{4}}{360\left(\bar{a}^{2}+\bar{b}^{2}\right)^{3}m^{4}\pi^{2}}$ & $\psi^{(4)}(\bar{a})\bar{a}^{7}+\bar{b}^{2}\psi^{(4)}(\bar{a})\bar{a}^{5}+3\bar{a}^{2}+15\bar{b}^{2}+24\left(\bar{a}^{5}+2\bar{b}^{2}\bar{a}^{3}\right)\psi^{(2)}(\bar{a})+\left(11\bar{a}^{6}+15\bar{b}^{2}\bar{a}^{4}\right)\psi^{(3)}(\bar{a})$\tabularnewline
\hline
\multirow{2}{*}{$\mathcal{L}_{SA}^{(1)}$} & \multirow{2}{*}{$-\frac{\bar{a}^{2}e^{4}\sigma}{360\bar{b}\left(\bar{a}^{2}+\bar{b}^{2}\right)^{3}m^{4}\pi^{2}}$} & $\psi^{(4)}(\bar{a})\bar{a}^{9}+\bar{b}^{2}\psi^{(4)}(\bar{a})\bar{a}^{7}-3\bar{a}^{4}-3\bar{b}^{2}\bar{a}^{2}-12\bar{b}^{4}+12\left(\bar{a}^{7}+\bar{b}^{2}\bar{a}^{5}-2\bar{b}^{4}\bar{a}^{3}\right)\psi^{(2)}(\bar{a})$\tabularnewline
 &  & $+\left(9\bar{a}^{8}+9\bar{b}^{2}\bar{a}^{6}-4\bar{b}^{4}\bar{a}^{4}\right)\psi^{(3)}(\bar{a})$\tabularnewline
\hline
\multirow{2}{*}{$\mathcal{L}_{AA}^{(1)}$} & \multirow{2}{*}{$-\frac{\bar{a}e^{4}\sigma^{2}}{360\bar{b}^{2}\left(\bar{a}^{2}+\bar{b}^{2}\right)^{3}m^{4}\pi^{2}}$} & $\psi^{(4)}(\bar{a})\bar{a}^{11}+\bar{b}^{2}\psi^{(4)}(\bar{a})\bar{a}^{9}+\left(6\bar{a}^{4}+\bar{b}^{2}\bar{a}^{2}-9\bar{b}^{4}\right)\psi^{(3)}(\bar{a})\bar{a}^{6}+9\bar{b}^{2}\bar{a}^{4}+9\bar{b}^{4}\bar{a}^{2}+12\bar{b}^{6}$\tabularnewline
 &  & $+6\left(\bar{a}^{9}-3\bar{b}^{4}\bar{a}^{5}+2\bar{b}^{6}\bar{a}^{3}\right)\psi^{(2)}(\bar{a})$\tabularnewline
\hline
\end{tabular}

\caption{Contribution to $\mathcal{L}_{S}^{(1)}$, $\mathcal{L}_{A}^{(1)}$,
$\mathcal{L}_{SS}^{(1)}$, $\mathcal{L}_{SA}^{(1)}$, and $\mathcal{L}_{AA}^{(1)}$
from the $O(\tilde{b}^{4})$ term in the expansion of $\mathcal{\bar{\mathcal{L}}}^{(1)}(\bar{a},\tilde{b})$
(\ref{L1ab_lowest3}). Each factor in the second column should be
multiplied to the corresponding formula in the third column. $\bar{b}=\bar{a}/\tilde{b}$. }

\label{tab:SAL1_n2}
\end{table}

\begin{table}
\begin{tabular}{|c|c|l|}
\hline
$\mathcal{L}_{S}^{(1)}$ & $-\frac{e^{2}}{24\pi^{2}}$ & $6\text{\ensuremath{\bar{a}}}^{2}-6\ln(2\pi)\bar{a}+12\ln(\Gamma(\bar{a}))\bar{a}+(2-6\bar{a})\ln(\bar{a})-24\zeta'(-1,\bar{a})+1$\tabularnewline
\hline
$\mathcal{L}_{SS}^{(1)}$ & $\frac{\bar{a}^{2}e^{4}}{3m^{4}\pi^{2}}$ & $6\psi^{(0)}(\bar{a})\bar{a}^{2}-6\bar{a}^{2}-3\ln(\bar{a})\bar{a}+3\ln(2\pi)\bar{a}-6\ln(\Gamma(\bar{a}))\bar{a}+3\bar{a}+1$\tabularnewline
\hline
$\mathcal{L}_{AA}^{(1)}$ & $-\frac{\bar{a}e^{4}\sigma^{2}}{6m^{4}\pi^{2}}$ & $6\bar{a}^{3}-6\ln(\bar{a})\bar{a}^{2}-6\ln(2\pi)\bar{a}^{2}+12\ln(\Gamma(\bar{a}))\bar{a}^{2}+2\psi^{(0)}(\bar{a})\bar{a}-24\zeta^{(1,0)}(-1,\bar{a})\bar{a}+\bar{a}+1$\tabularnewline
\hline
\end{tabular}

\caption{$\mathcal{L}_{S}^{(1)}$, $\mathcal{L}_{SS}^{(1)}$, and $\mathcal{L}_{AA}^{(1)}$
for $\tilde{b}=0$. $\mathcal{L}_{A}^{(1)}$ and $\mathcal{L}_{SA}^{(1)}$
are null. Each factor in the second column should be multiplied to
the corresponding formula in the third column. $\bar{b}=\bar{a}/\tilde{b}$. }

\label{tab:SAL1_n1binf}
\end{table}

The weak-field limit of the quantities in (\ref{SA_L1}) can be obtained
by combining the results in Tabs.~\ref{tab:SAL1_n0}, \ref{tab:SAL1_n1},
and \ref{tab:SAL1_n2}, and finding the asymptotic form for the limit
of $\bar{a},\bar{b}\rightarrow\infty$. Alternatively, the weak-field
limit of the effective Lagrangian (\ref{L1ab_weakO6}) can be used
to evaluate $\mathcal{L}_{S}^{(1)}$, $\mathcal{L}_{A}^{(1)}$, $\mathcal{L}_{SS}^{(1)}$,
$\mathcal{L}_{SA}^{(1)}$, and $\mathcal{L}_{AA}^{(1)}$:

\begin{equation}
\mathcal{L}_{S}^{(1)}=\frac{e^{2}}{360\pi^{2}}\left(\frac{1}{\bar{a}^{2}}-\frac{1}{\bar{b}^{2}}\right),\quad\mathcal{L}_{A}^{(1)}=\frac{7e^{2}\sigma}{720\text{\ensuremath{\bar{a}}}\bar{b}\pi^{2}},
\end{equation}
\begin{equation}
\mathcal{L}_{SS}^{(1)}=\frac{e^{4}}{45m^{4}\pi^{2}},\quad\mathcal{L}_{SA}^{(1)}=0,\quad\mathcal{L}_{AA}^{(1)}=\frac{7e^{4}}{180m^{4}\pi^{2}},
\end{equation}
which are obtained with the quantities of the first bracket in (\ref{L1ab_weakO6})
only. When $\bar{b}=\infty$, these expressions yields the permittivity
and permeability tensors obtained by Adler for the case of a weak
magnetic field \citep{Adler1971Photon}. If the quantities of the
second bracket in (\ref{L1ab_weakO6}) is used, higher-order contributions
are obtained.

\section{Vacuum birefringence for $\mathbf{B}_{0}\parallel\mathbf{E}_{0}$
and $|\mathbf{B}_{0}|\gg|\mathbf{E}_{0}|$\label{sec:VacBi}}

\begin{figure}
\includegraphics[width=0.2\textwidth]{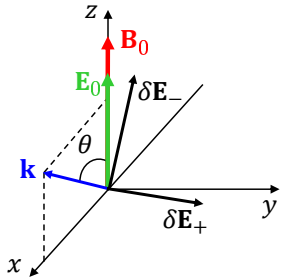}

\caption{Configuration of the background fields ($\mathbf{E}_{0}$, $\mathbf{B}_{0}$)
and the probe field. The probe field has its propagation vector $\mathbf{k}$
on the $xz$-plane and its two polarization vectors $\delta\mathbf{E}_{\pm}$
associated with the refractive indices $n_{\pm}$. Unless $E_{0}=0$
(thus $\epsilon_{2}=0$), $\delta\mathbf{E}_{+}$ is not along the
$y$-axis, and $\delta\mathbf{E}_{-}$ is not on the $xz$-plane.}

\label{fig:config}
\end{figure}

In this section, we work out the refractive indices and the polarization
vectors for the case of a weak electric field added parallel to an
arbitrarily strong magnetic field, as shown in Fig.~\ref{fig:config}.
In such a configuration, $a=B_{0}$, $b=E_{0}$, and $\bar{a}=B_{c}/(2B_{0})$.
This configuration, looking too restrictive at a first glance, is
actually general enough to include the non-parallel cases, too. By
choosing an appropriate Lorentz transformation, one can transform
non-perpendicular configurations into parallel ones and the perpendicular
configuration into that of a pure magnetic one as far as the electric
field is weaker than the magnetic field \citep{Jackson1999Classical,Melrose2013Quantum}.
However, the Lorentz transformation of the permittivity and permeability
of anisotropic media is a highly non-trivial issue \citep{ODell1962electrodynamics}.
Furthermore, for the localized fields of the pulsar magnetosphere,
such a parallelization of fields by a Lorentz transformation can be
done only locally. We leave the resolution of these issues as future
works.

The refractive indices and the associated polarization vectors are
found by solving the Maxwell equations for the probe field. When the
probe field is a plane wave with the propagation vector $\mathbf{k}$
and the angular frequency $\omega$ ($\mathbf{k}=\omega\mathbf{n}=\omega n\hat{k}$),
the Maxwell equations for the probe field reduce to

\begin{equation}
\omega\delta\mathbf{B}=\mathbf{k}\times\delta\mathbf{E},\quad\omega\delta\mathbf{D}=-\mathbf{k}\times\delta\mathbf{H}.
\end{equation}
By substituting (\ref{dDdH}) into these equations, we obtain a matrix-vector
equation:

\begin{equation}
\boldsymbol{\epsilon}_{E}\cdot\delta\mathbf{E}+\boldsymbol{\epsilon}_{B}\cdot\mathbf{n}\times\delta\mathbf{E}+\mathbf{n}\times\left(\bar{\boldsymbol{\mu}}_{B}\cdot\mathbf{n}\times\delta\mathbf{E}+\bar{\boldsymbol{\mu}}_{E}\cdot\delta\mathbf{E}\right)=\boldsymbol{\Lambda}\cdot\delta\mathbf{E}=0,
\end{equation}
where $\boldsymbol{\Lambda}$ is a $3\times3$ matrix incorporating
$\boldsymbol{\epsilon}_{E}$, $\boldsymbol{\epsilon}_{B}$, $\bar{\boldsymbol{\mu}}_{B}$,
$\bar{\boldsymbol{\mu}}_{E}$, and $\mathbf{n}$. For this equation
to have non-trivial solutions, $\mathrm{det}\boldsymbol{\Lambda}=0$
should hold, from which refractive indices ($n_{\pm}$) are obtained.
Substituting each of the refractive indices into the matrix-vector
equation, one can obtain the associated polarization vectors ($\delta\mathbf{E}_{\pm}$).

\begin{table}
\begin{tabular}{|c|c|c|}
\hline
$\boldsymbol{r}$ & $r$ & $\tilde{r}$\tabularnewline
\hline
\hline
$\boldsymbol{\epsilon}_{E}$ & $\epsilon_{E}=1-\mathcal{L}_{S}^{(1)}$ & $\tilde{\epsilon}_{E}=E_{0}^{2}\mathcal{L}_{SS}^{(1)}+2E_{0}B_{0}\mathcal{L}_{SA}^{(1)}+B_{0}^{2}\mathcal{L}_{AA}^{(1)}$\tabularnewline
\hline
$\boldsymbol{\epsilon}_{B}$ & $\epsilon_{B}=-\mathcal{L}_{A}^{(1)}$ & $\tilde{\epsilon}_{B}=-E_{0}B_{0}\mathcal{L}_{SS}^{(1)}+\left(E_{0}^{2}-B_{0}^{2}\right)\mathcal{L}_{SA}^{(1)}+B_{0}E_{0}\mathcal{L}_{AA}^{(1)}$\tabularnewline
\hline
$\bar{\boldsymbol{\mu}}_{B}$ & $\bar{\mu}_{B}=1-\mathcal{L}_{S}^{(1)}$ & $\tilde{\bar{\mu}}_{B}=-B_{0}^{2}\mathcal{L}_{SS}^{(1)}+2E_{0}B_{0}\mathcal{L}_{SA}^{(1)}-E_{0}^{2}\mathcal{L}_{AA}^{(1)}$\tabularnewline
\hline
$\bar{\boldsymbol{\mu}}_{E}$ & $\bar{\mu}_{E}=\mathcal{L}_{A}^{(1)}$ & $\tilde{\bar{\mu}}_{E}=E_{0}B_{0}\mathcal{L}_{SS}^{(1)}-\left(E_{0}^{2}-B_{0}^{2}\right)\mathcal{L}_{SA}^{(1)}-B_{0}E_{0}\mathcal{L}_{AA}^{(1)}$\tabularnewline
\hline
\end{tabular}

\caption{Components of the permittivity and permeability tensors for the configuration
in Fig.~\ref{fig:config}. These tensors have the shape of (\ref{tensor_structure}).}

\label{tab:emu_comps}
\end{table}

In the configuration in Fig.~\ref{fig:config}, the parallel field
condition $\mathbf{B}_{0}=B_{0}\hat{z}$ and $\mathbf{E}_{0}=E_{0}\hat{z}$
($E_{0}\ge0$ and $B_{0}>0$) forces the permittivity and permeability
tensors $\boldsymbol{\epsilon}_{E}$, $\boldsymbol{\epsilon}_{B}$,
$\bar{\boldsymbol{\mu}}_{B}$, and $\bar{\boldsymbol{\mu}}_{E}$ in
(\ref{dDdH}) to have the following structure:

\begin{equation}
\boldsymbol{r}=\left(\begin{array}{ccc}
r & 0 & 0\\
0 & r & 0\\
0 & 0 & r+\tilde{r}
\end{array}\right),\label{tensor_structure}
\end{equation}
where $r$ and $\tilde{r}$ are given in Tab.~\ref{tab:emu_comps}
for each tensor. Furthermore, without loss of generality, the propagation
vector can be assumed to be in the $xz$-plane: $\mathbf{k}=\omega n(\sin\theta,0,\cos\theta)$.
Then the matrix $\boldsymbol{\Lambda}$ becomes

\begin{equation}
\boldsymbol{\Lambda}=\left(\begin{array}{ccc}
1-n^{2}\cos^{2}\theta & 0 & n^{2}\sin\theta\cos\theta\\
0 & 1-n^{2}(1+\mu\sin^{2}\theta) & \epsilon_{2}n\sin\theta\\
n^{2}\sin\theta\cos\theta & \epsilon_{2}n\sin\theta & 1+\epsilon_{1}-n^{2}\sin^{2}\theta
\end{array}\right),\label{Lamda_reduced}
\end{equation}
where

\begin{equation}
\mu=\frac{\tilde{\bar{\mu}}_{B}}{\epsilon_{E}},\quad\epsilon_{1}=\frac{\tilde{\epsilon}_{E}}{\epsilon_{E}},\quad\epsilon_{2}=\frac{\tilde{\epsilon}_{B}}{\epsilon_{E}}.
\end{equation}
Below we assume $(\epsilon_{1}+\mu+\epsilon_{1}\mu)<0$ that held
in all the cases we studied. When $(\epsilon_{1}+\mu+\epsilon_{1}\mu)>0$,
the two modes denoted by the subscript $\pm$ are swapped in the wrenchless
case.

Solving the matrix equation is straightforward, but the general results
are too lengthy to be presented. Instead, noting that $\epsilon_{2}$
vanishes with $E_{0}$ (In Tab.~\ref{tab:emu_comps}, $\mathcal{L}_{SA}^{(1)}=0$
when $E_{0}=0$), we present only the expansions of the refractive
indices and the polarization vectors up to $O(\epsilon_{2}^{2})$
below. From $\mathrm{det}\boldsymbol{\Lambda}=0$, the refractive
indices are obtained obtained:

\begin{equation}
n_{+}^{2}\simeq\frac{1}{1+\mu\sin^{2}\theta}-\frac{(1+\mu)\sin^{2}\theta}{(\epsilon_{1}+\mu+\epsilon_{1}\mu)(1+\mu\sin^{2}\theta)^{2}}\epsilon_{2}^{2},\label{n2p}
\end{equation}

\begin{equation}
n_{-}^{2}\simeq\frac{1+\epsilon_{1}}{1+\epsilon_{1}\cos^{2}\theta}+\frac{(1+\epsilon_{1})\sin^{2}\theta}{(\epsilon_{1}+\mu+\epsilon_{1}\mu)(1+\epsilon_{1}\cos^{2}\theta)^{2}}\epsilon_{2}^{2}.\label{n2n}
\end{equation}

Then, the polarization vectors corresponding to $n_{\pm}^{2}$ are
obtained by solving $\boldsymbol{\Lambda}_{\pm}\cdot\delta\mathbf{E}_{\pm}=0$,
where $\boldsymbol{\Lambda}_{\pm}$ is $\boldsymbol{\Lambda}$ (\ref{Lamda_reduced})
with $n^{2}=n_{\pm}^{2}$. However, a blind application of the Gauss
elimination fails to find the correct solution that becomes the wrenchless
solution as $\epsilon_{2}$ vanishes. By considering the limiting
behavior of $\Lambda_{ij}$ for $\epsilon_{2}\rightarrow0$ and requiring
the solution's limiting behavior be consistent with that of $\Lambda_{ij}$,
one can make $\boldsymbol{\Lambda}$, $\delta\mathbf{E}$, and $\boldsymbol{\Lambda}\cdot\delta\mathbf{E}=0$
safely reduce those of the wrenchless case as $\epsilon_{2}\rightarrow0$.
For example, as $\epsilon_{2}\rightarrow0$, $\Lambda_{+,22}\rightarrow\epsilon_{2}^{2}$,
and $\Lambda_{+,23}\rightarrow\epsilon_{2}$, while other non-zero
components do not vanish. Then, the solution of the type $\left(O(\epsilon_{2}),1,O(\epsilon_{2})\right)^{T}$
yields the correct solution in terms of $\Lambda_{ij}$:

\begin{equation}
\delta\mathbf{E}_{+}=\left(\begin{array}{c}
\frac{\Lambda_{13}\Lambda_{23}}{\Lambda_{11}\Lambda_{33}-\Lambda_{13}^{2}}\\
1\\
-\frac{\Lambda_{11}\Lambda_{23}}{\Lambda_{11}\Lambda_{33}-\Lambda_{13}^{2}}
\end{array}\right)_{n^{2}=n_{+}^{2}}\simeq\left(\begin{array}{c}
0\\
1\\
0
\end{array}\right)+\left(\begin{array}{c}
\frac{1}{\epsilon_{1}+\mu+\epsilon_{1}\mu}\frac{\cos\theta}{\sqrt{1+\mu\sin^{2}\theta}}\\
0\\
-\frac{1}{\epsilon_{1}+\mu+\epsilon_{1}\mu}\frac{(1+\mu)\sin\theta}{\sqrt{1+\mu\sin^{2}\theta}}
\end{array}\right)\epsilon_{2},\label{dEp}
\end{equation}
where the denominator $\Lambda_{11}\Lambda_{33}-\Lambda_{13}^{2}$
does not vanish as $\epsilon_{2}\rightarrow0$. Similarly, as $\epsilon_{2}\rightarrow0$,
$\Lambda_{-,23}\rightarrow\epsilon_{2}$, $\Lambda_{-,33}-\Lambda_{-,13}^{2}/\Lambda_{-,11}\rightarrow\epsilon_{2}^{2}$,
while other non-zero components do not vanish. Then the solution of
the type $\left(x,O(\epsilon_{2}),1\right)^{T}$, where $x$ is non-vanishing,
yields the following solution:

\begin{equation}
\delta\mathbf{E}_{-}=\left(\begin{array}{c}
-\frac{\Lambda_{13}}{\Lambda_{11}}\\
-\frac{\Lambda_{23}}{\Lambda_{22}}\\
1
\end{array}\right)_{n^{2}=n_{-}^{2}}\simeq\left(\begin{array}{c}
-(1+\epsilon_{1})\cot\theta\\
0\\
1
\end{array}\right)+\left(\begin{array}{c}
0\\
\frac{\sqrt{1+\epsilon_{1}}}{\epsilon_{1}+\mu+\epsilon_{1}\mu}\frac{\sqrt{1+\epsilon_{1}\cos^{2}\theta}}{\sin\theta}\\
0
\end{array}\right)\epsilon_{2}+\left(\begin{array}{c}
-\frac{(1+\epsilon_{1})\cot\theta}{\epsilon_{1}+\mu+\epsilon_{1}\mu}\\
0\\
0
\end{array}\right)\epsilon_{2}^{2},\label{dEn}
\end{equation}
where $\Lambda_{11}$ and $\Lambda_{22}$ do not vanish as $\epsilon_{2}\rightarrow0$.

In the wrenchless case, $E_{0}=0$ and, thus, $\epsilon_{2}=0$ hold.
The corresponding refractive indices and polarization vectors are
the zeroth-order terms in (\ref{n2p}), (\ref{n2n}), (\ref{n2n}),
and (\ref{dEn}), which are consistent with those obtained by Melrose
\citep{Melrose2013Quantum}. Unlike in the background-field-free vacuum,
$\mathbf{k}$, $\delta\mathbf{E}_{+}$, and $\delta\mathbf{E}_{-}$
do not form an orthogonal triad in general, albeit $\mathbf{k}\perp\delta\mathbf{E}_{+}$
and $\delta\mathbf{E}_{+}\perp\delta\mathbf{E}_{-}$. The polarization
vector $\delta\mathbf{E}_{+}$ is along the $y$-axis, while $\delta\mathbf{E}_{-}$
lies in the $xz$-plane but not necessarily $\mathbf{k}\perp\delta\mathbf{E}_{-}$
\citep{Hu2007Birefringence}. An orthogonal triad is formed only when
$\theta=\pi/2$: $\delta\mathbf{E}_{-}$ is aligned along the $z$-axis.

When $\theta=0$, i.e., $\mathbf{k}$ is along the $z$-axis, $n_{\pm}^{2}=1$
and $\delta\mathbf{E}_{\pm,z}=0$ regardless of the wrench: the background
field does not affect the propagation of the probe field because of
the equal and opposite contributions from the virtual electrons and
positrons.

\begin{figure}
\includegraphics[width=1\textwidth]{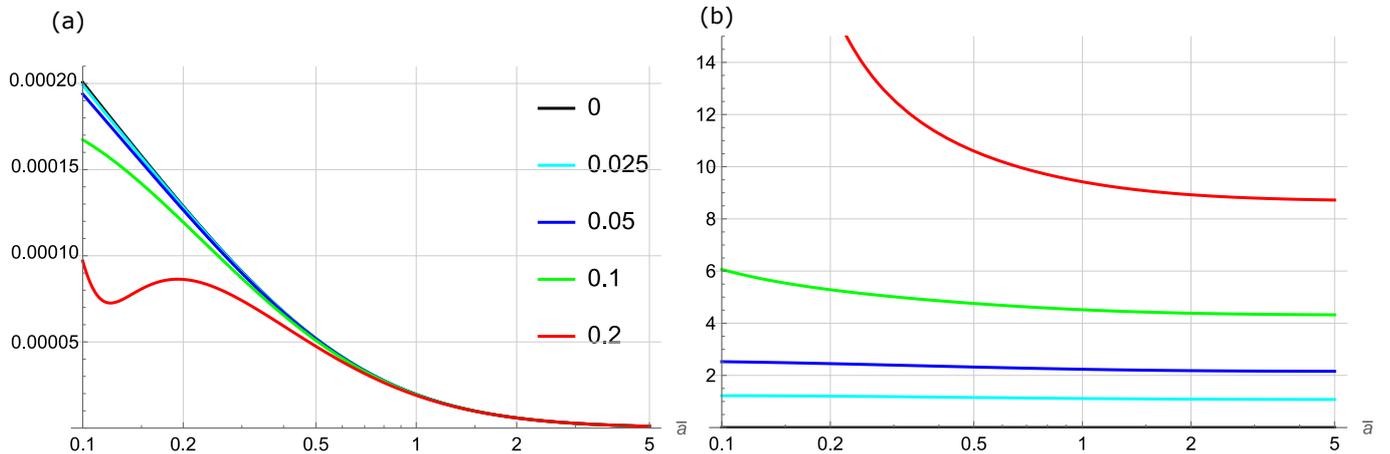}

\caption{Effect of the electromagnetic wrench on vacuum birefringence: (a)
$n_{+}-n_{-}$ and (b) the angle (degree) of $\delta\mathbf{E}_{+}$
with respect to the $y$-axis for $\tilde{b}=0,0.025,0.05,0.1,0.2$.
The propagation vector $\mathbf{k}$ of the probe field is along the
$x$-axis ($\theta=\pi/2$). }

\label{fig:dn_ang}
\end{figure}

The electromagnetic wrench, i.e., non-zero $\tilde{b}$, can significantly
affect the vacuum birefringence, as shown in Fig.~\ref{fig:dn_ang}.
In Fig.~\ref{fig:dn_ang}(a), the difference of the two refractive
indices, $n_{+}-n_{-}$, decreases noticeably for small $\bar{a}$
as $\tilde{b}$ increases: a 15\% of reduction at $\bar{a}=0.1$ for
$\tilde{b}=0.1$ ($B=B_{\mathrm{cr}}$ and $E=E_{\mathrm{cr}}$) albeit
the reduction is negligible in the subcritical region. For $\tilde{b}=0.2$,
the reduction is about 50\% in the supercritically magnetic region,
in which pair production is not negligible for $\tilde{b}=0.2$. In
addition, the polarization vectors rotates due to the wrench, as shown
in Fig.~\ref{fig:dn_ang}(b). When $\mathbf{k}$ is along the $x$-axis
($\theta=\pi/2$), $\delta\mathbf{E}_{+}$ and $\delta\mathbf{E}_{-}$
on the $yz$-plane, and they are almost perpendicular to each other.
In Fig.~\ref{fig:dn_ang}(b), $\delta\mathbf{E}_{+}$ rotates from
the $y$-axis to the $z$-axis as $\tilde{b}$ increases. The increase
does not depend on $\bar{a}$ until $\tilde{b}$ reaches 0.1. For
$\tilde{b}=0.2,$the rotation angle is about 60 degrees at $\bar{a}=0.1$.
Even at lower values of $\tilde{b}$, the rotation angle amounts to
a few degrees. In contrast, the wrenchless case allows no such rotation
when $\theta=\pi$/2. Both the reduction of the differences of the
refractive indices and the rotation of polarization vectors are new
features of vacuum birefringence introduced by the non-zero electromagnetic
wrench. These results suggest that the electromagnetic wrench should
be taken into account in analyzing the vacuum birefringence when the
electric field is a fraction of a supercritical magnetic field. Such
a situation is anticipated in the magnetospheres of highly magnetized
pulsars and neutron stars.

\section{Conclusion\label{sec:Conclusion}}

We have provided an explicit closed expression of the one-loop effective
Lagrangian for the vacuum under an arbitrarily strong magnetic field
superposed with a weaker electric field. To our knowledge, previous
studies focused on the wrenchless case of $G=-\mathbf{E}\cdot\mathbf{B}=0$.
But our expression is valid for the cases of $G\neq0$ as far as the
pair production is not significant; in the case of significant pair
production, a plasma of produced electron-positron pairs affects the
vacuum polarization. Furthermore, the provided expressions, (\ref{L1ab_series})
and (\ref{L1ab_lowest3}), is compact to facilitate theoretical analysis.

From the explicit closed expression, we have calculated the linear
optical response of such vacuum to weak low-frequency fields. The
permittivity and permeability tensors are given as (\ref{eEij}),
(\ref{eBij}), (\ref{muBij}), and (\ref{muBij}) for an arbitrary
one-loop effective Lagrangian. When the expansion form (\ref{L1ab_lowest3})
of the Heisenberg\textendash Euler and Schwinger effective Lagrangian
is used, these tensors have values specified in Tabs.~\ref{tab:SAL1_n0},
\ref{tab:SAL1_n1}, and \ref{tab:SAL1_n2}. The known results for
the wrenchless and the weak-field cases in the literature are obtained
by taking the limit of $b\rightarrow0$ for arbitrary $a$ and $a,b\rightarrow0$
in our general expression, respectively.

With the permittivity and permeability tensors, we have worked out
the modes of the probe field for the case where the background electric
and magnetic fields are parallel to each other. The two refractive
indices (\ref{n2p}) and (\ref{n2n}) clearly exhibits birefringence
with the associated polarization vectors (\ref{dEp}) and (\ref{dEn}).
Similarly, the results of the wrenchless case in the literature are
the above limit of our expression. In the case with electromagnetic
wrench, we have showed that electromagnetic wrench can reduce the
difference of the refractive indices and rotate the polarization tensors;
these effects have not been reported so far to our knowledge.

Our results can be used for the x-ray polarimetry of highly magnetized
neutron stars. In the magnetosphere of such astrophysical
objects, magnetic fields are comparable to or higher than the Schwinger
field and accompanied by the induced weak electric fields. The electric
field can noticeably change the polarimetric results, as shown in
Fig.~\ref{fig:dn_ang}. For instance, when $B=5B_{\mathrm{cr}}$
and $E=0.5E_{\mathrm{cr}}$, the difference of the refractive indices
are changed by 15\%, and the polarization vectors rotate by 6$^{\circ}$
due to the non-negligible electric field along the magnetic field.
At smaller values of $E$, the change is reduced but can grow to a
significant level because the probe's propagation length is comparable
to the size of the stars. Our expression that takes electromagnetic
wrench into account enables an accurate analysis for such conditions.
\begin{acknowledgments}
This work was supported by Institute for Basic Science (IBS) under
IBS-R012-D1. The work of SPK was also supported by National Research
Foundation of Korea (NRF) funded by the Ministry of Education (2019R1I1A3A01063183).
\end{acknowledgments}

\appendix

\section{Numerical evaluation of the integral expression of $\mathcal{L}^{(1)}(a,b)$\label{sec:Numerical}}

We consider the numerical evaluation of the integral in (\ref{L1ab_int}).
The integrand has a proper singularity at $z=0$ with a well-defined
limit (0) and thus poses no problem in principle. In numerical evaluation,
however, the functional form in (\ref{L1ab_int_z}) leads to a serious
loss of significant digits near $z=0$. This problem can be avoided
by using the Taylor expansion of the integrand around $z=0$. More
problematic are the poles of the $\cot(\tilde{b}z)$ at $z=n\pi/\tilde{b}$
$(n=1,2,\dots)$. As we are interested only in the real part of $\mathcal{L}^{(1)}(a,b)$
to study vacuum birefringence, we take the principal value of the
integral.

Taking into account of these problems, we split the integral into
two parts, $I=I_{1}+I_{2}$: one from 0 to $z_{b}$ ($z_{b}\sim0$
and $z_{b}\ll\pi/\tilde{b}$, the first pole) and the other from $z_{b}$
to $(n+1/2)\pi/\tilde{b}$, the midpoint between the $n$-th pole
and $(n+1)$-th pole. As the number of poles increases, the numerical
integration would converge to the exact value of the integral in (\ref{L1ab_int_z}).
In the first part, we use the second-order Taylor expansion:

\begin{equation}
I_{1}(\bar{a},\tilde{b},n)=\int_{0}^{z_{b}}e^{-2\bar{a}z}\frac{z}{945}\left[21(1+5\tilde{b}^{2}+\tilde{b}^{4})+z^{2}(-2-7\tilde{b}^{2}+7\tilde{b}^{4}+2\tilde{b}^{6})\right]\mathrm{d}z.
\end{equation}
The second part is given as

\begin{equation}
I_{2}(\bar{a},\tilde{b},n)=\mathrm{Pr}\int_{z_{b}}^{\frac{(n+1/2)\pi}{\tilde{b}}}\frac{e^{-2\bar{a}z}}{z^{3}}\left[1+\frac{z^{2}(1-\tilde{b}^{2})}{3}-\tilde{b}z^{2}\coth(z)\cot(z\tilde{b})\right]\mathrm{d}z,
\end{equation}
where $\mathrm{Pr}$ means the Cauchy principal value. To numerically
evaluate the Cauchy principal value, we used Mathematica \citep{Mathematica},
which implements the algorithm presented in 2.12.8 of \citep{Davis2014Methods}.
In the parameter range considered in our study, $z_{b}=0.01$ and
$n\le20$ gave a good convergence. A small number of poles are sufficient
for convergence as a small value of $\tilde{b}$ pushes the poles
away from the origin, and the contribution from the region far from
the origin is significantly suppressed by the factor $e^{-2\bar{a}z}/z^{3}$.

\section{Analytic expression of \textmd{\textup{\normalsize{}$\mathcal{L}^{(1)}(a,0)$}}
\label{app:La0}}

The integral expression of the one-loop effective action with $b=0$
is obtained by taking the limit of $\tilde{b}=0$ and using $\lim_{x\rightarrow0}x\cot(x)=1$
in (\ref{L1ab_int_z}):

\begin{equation}
\mathcal{\bar{L}}^{(1)}(a,0)=\frac{m^{4}}{8\pi^{2}}\frac{1}{4\bar{a}^{2}}\int_{0}^{\infty}\frac{e^{-2\bar{a}z}}{z^{3}}\left[1+\frac{z^{2}}{3}-z\coth(z)\right]\mathrm{d}z,\label{L1a_int}
\end{equation}
where $\bar{a}=m^{2}/(2ea)$. The integration can be performed by
expanding $z\coth z$ is expanded around $z=0$:

\begin{equation}
z\coth z=\sum_{n=0}^{\infty}\frac{B_{2n}(2z)^{2n}}{(2n)!},\label{zcothz}
\end{equation}
where $B_{2n}$ are the Bernoulli numbers. This expansion is convergent
for $|z|<\pi$, as can be seen by the root test. However, it can be
substituted into (\ref{L1a_int}) to yield an asymptotic expression
for $\bar{a}\rightarrow\infty$ because $\exp\left(-2\bar{a}z\right)$
suppresses the contribution from the region of $z\gg1/\bar{(2a)}$
if the remaining part of the integrand has a polynomial divergence
at most. By using the formula $\int_{0}^{\infty}e^{-\alpha z}z^{p}\,dz=\Gamma(p+1)/\alpha^{p+1}$,
we can obtain an asymptotic expression of $\mathcal{L}^{(1)}(a,0)$:

\begin{equation}
\mathcal{\bar{L}}^{(1)}(a,0)\sim-\frac{m^{4}}{8\pi^{2}}\sum_{n=2}^{\infty}\frac{B_{2n}}{2n(2n-1)(2n-2)}\frac{1}{\bar{a}^{2n}}.\label{L1a_series}
\end{equation}
This series (\ref{L1a_series}) is divergent for any finite value
of $\bar{a}$, which can be found by the root test, and, as $\bar{a}\rightarrow\infty$,
it is asymptotic to a function involving the Hurwitz zeta function
$\zeta(s,\bar{a})$ (25.11.44 in \citep{Olver2010NIST}):

\begin{equation}
-\sum_{n=2}^{\infty}\frac{B_{2n}}{2n(2n-1)(2n-2)}\frac{1}{\bar{a}^{2n-2}}\sim H(\bar{a})=\zeta'(-1,\bar{a})-\frac{1}{12}+\frac{\bar{a}^{2}}{4}-\left(\frac{1}{12}-\frac{\bar{a}}{2}+\frac{\bar{a}^{2}}{2}\right)\ln\bar{a},\label{Ha}
\end{equation}
where $\zeta'(-1,\bar{a})=d\zeta(s,\bar{a})/ds|_{s=-1}$. Consequently,

\begin{equation}
\bar{\mathcal{L}}^{(1)}(\bar{a},0)\sim\frac{m^{4}}{8\pi^{2}}\frac{H(\bar{a})}{\bar{a}^{2}}.\label{L1a_Ha}
\end{equation}
Remarkably, this asymptotic relation turns out to be equality. The
formula (\ref{L1a_Ha}) is exactly the expression of $\mathcal{\bar{L}}^{(1)}(\bar{a},0)$
obtained either by the dimensional regularization of (\ref{L1a_int})
\citep{Dittrich1976One,Dittrich1979Evaluation} or by the Schwinger-DeWitt
in-out formalism with $\Gamma$-function regularization \citep{Kim2019Quantum}.

We mention a useful symmetry of the effective Lagrangian. From (\ref{FG_EB}),
$F$ and $|G|$ are invariant when $a$ ($b$) is replaced by $ib$
($-ia$), leading to a symmetry relation:

\begin{equation}
\mathcal{L}(a,b)=\mathcal{L}(ib,-ia).
\end{equation}
Therefore, when the expression of $\mathcal{L}(a,0)$ is available,
setting $a=ib'$ yields the expression of $\mathcal{L}(0,b')$:

\begin{equation}
\mathcal{L}(a,0)=\mathcal{L}(ib',0)=\mathcal{L}(0,-i(ib'))=\mathcal{L}(0,b')=\frac{m^{4}}{8\pi^{2}}\frac{1}{\bar{b'}^{2}}\left[\frac{\pi}{4\alpha}-H(-i\bar{b'})\right],\label{La0}
\end{equation}
which matches the one-loop effective Lagrangian in a uniform electric
field. Further, note that the imaginary part of (\ref{La0}) is equivalent
to the sum of the residues from the simple poles of (\ref{L1ab_int}),
which was explicitly shown in \citep{Kim2019Quantum}.

\section{Expression of $H^{(2n)}(z)$\label{App:H2nz}}

The even-order derivatives of $H(z)$ can be explicitly obtained.
The function $H(z)$ consists of two parts:

\begin{equation}
H(z)=\zeta'(-1,z)+h(z)=\zeta'(-1,z)-\frac{1}{12}+\frac{z^{2}}{4}-\left(\frac{1}{12}-\frac{z}{2}+\frac{z^{2}}{2}\right)\ln z,\label{hz}
\end{equation}
where $\zeta'(-1,z)=d\zeta(s,z)/ds|_{s=-1}$. The function $\zeta(s,z)$
is the Hurwitz zeta function, defined as (25.11.1 in \citep{Olver2010NIST}):

\[
\zeta\left(s,z\right)=\sum_{n=0}^{\infty}\frac{1}{(n+z)^{s}}\quad(\mathrm{Re}\{s\}>1,\;z\neq0,-1,-2,\dots).
\]
As far as $s\neq1$, the expression can be analytically continued.
To calculate $\partial_{z}^{2n}\zeta'(-1,z)$, we begin with the following
identity (25.11.17 in \citep{Olver2010NIST}):

\begin{equation}
\partial_{z}\zeta(s,z)=-s\,\zeta(s+1,z),\quad(s\neq0,1\;\mathrm{and}\;\mathrm{Re}\{z\}>0).
\end{equation}
Differentiating with respect to $s$ and setting $s=-1$, we obtain

\begin{equation}
\partial_{z}\zeta'(-1,z)=-\zeta(0,z)+\zeta'(0,z)=z-\frac{1}{2}+\ln\Gamma(z)-\frac{1}{2}\ln(2\pi),
\end{equation}
where $\zeta(0,z)=-z+1/2$ and $\zeta'(0,z)=\ln\Gamma(z)-\ln(2\pi)/2$
are used for the second equality (25.11.13 and 25.11.18 in \citep{Olver2010NIST}).
Differentiating with respect to $z$ successively and using the definition
of the polygamma function (5.2.2 and 5.15 in \citep{Olver2010NIST})

\begin{equation}
\psi^{(m)}(z)=\mathrm{d}^{m+1}\left(\ln\Gamma(z)\right)/\mathrm{d}z^{m+1},
\end{equation}
we obtain the formula of $\partial_{z}^{2n}\zeta'(-1,z)$:

\begin{equation}
\partial_{z}^{2n}\zeta'(-1,z)=\delta_{n1}+\psi^{(2n-2)}(z),\quad n\ge1.
\end{equation}
The successive differentiation of $h(z)$ in (\ref{hz}) is straightforward,
and, consequently, the formula of $H^{(2n)}(z)$ is given as

\begin{equation}
H^{(2n)}(z)=\psi^{(2n-2)}(z)+\frac{1}{12}\frac{\Gamma(2n)}{z^{2n}}+\frac{1}{2}\frac{\Gamma(2n-1)}{z^{2n-1}}+\frac{\Gamma(2n-2)}{z^{2n-2}}\theta(n-2)-\delta_{n1}\ln z,\quad(n\ge1),\label{H2nz}
\end{equation}
where $\theta(n)$ is the unit step function with $\theta(n\ge0)=1$.

\bibliographystyle{apsrev4-2}
\bibliography{ref_VBir}

\end{document}